\documentclass[11pt,]{article}
\usepackage{authblk}
\usepackage{fullpage}
\usepackage{amssymb,amsmath}
\usepackage{soul}
\usepackage[T1]{fontenc}
\usepackage{siunitx}
\usepackage[version=3]{mhchem}
\usepackage{xr}
\usepackage{bm,bbm}
\usepackage{pdfpages}
\usepackage{float}
\usepackage{graphicx}
\usepackage{dcolumn}
\usepackage{bm}
\usepackage{color}
\usepackage{tikz}
\usepackage{setspace}
\usepackage{cancel}
\usepackage[title]{appendix}
\usepackage[center,font=small,labelfont=bf]{caption}
\usepackage{float}
\usepackage{braket}

\usepackage[unicode=true]{hyperref}
\hypersetup{colorlinks=true,
            citecolor=black,
            linkcolor=black}

\usepackage[authoryear,round]{natbib}

\usepackage{setspace}

\makeatletter

\newcommand{\hypgeo}[2]{%
  {\vphantom{F}}_{#1}\kern-\scriptspace F_{#2}%
}

\newcommand\approxsim{\mathchoice
  {\@approxsim {\displaystyle}      {1ex} }
  {\@approxsim {\textstyle}         {1ex} }
  {\@approxsim {\scriptstyle}       {.7ex}}
  {\@approxsim {\scriptscriptstyle} {.5ex}}}
\newcommand\@approxsim[2]{%
  \mathrel{%
    \ooalign{%
      $\m@th#1\sim$\cr
      \hidewidth$\m@th#1.$\hidewidth\cr
      \hidewidth\raise #2 \hbox{$\m@th#1.$}\hidewidth\cr
    }%
  }%
}

\newcommand{\bo}{\raise-1mm\hbox{\Large$\Box$}}

\begin{document}
\title{How animal movement influences wildlife-vehicle collision risk: a mathematical framework for range-resident species}

\author[1,2]{Benjamin Garcia de Figueiredo}
\author[3]{Inês Silva}
\author[4,5,6]{Michael J. Noonan}
\author[7]{Christen H. Fleming}
\author[8]{William F. Fagan}
\author[3,8,9,10]{Justin M. Calabrese}
\author[3,1,11*]{Ricardo Martinez-Garcia}
\linespread{1}

\affil[1]{\small ICTP South American Institute for Fundamental Research \& Instituto de F\'isica Te\'orica, Universidade
Estadual Paulista - UNESP, Brazil.}
\affil[2]{Joseph Henry Laboratories of Physics and Lewis–Sigler Institute for Integrative Genomics,
Princeton University, USA}
\affil[3]{Center for Advanced Systems Understanding (CASUS) -- Helmholtz-Zentrum Dresden-Rossendorf (HZDR), Germany}
\affil[4]{Department of Biology, The University of British Columbia Okanagan, Kelowna, V1V 1V7, Canada.}
\affil[5]{Okanagan Institute for Biodiversity, Resilience, and Ecosystem Services, The University of British Columbia Okanagan, Kelowna, V1V 1V7, Canada.}
\affil[6]{Department of Computer Science, Math, Physics, and Statistics, The University of British Columbia Okanagan, Kelowna, V1V 1V7, Canada.}
\affil[7]{Department of Biology, University of Central Florida, Biological Sciences Bldg. 4110 Libra Drive, Orlando, Florida 32816-2368}
\affil[8]{Department of Biology, University of Maryland, College Park, MD 20742, USA}
\affil[9]{Department of Ecological Modelling, Helmholtz Centre for Environmental Research -- UFZ, Germany}
\affil[10]{Department of Geosciences, TUD Dresden University of Technology, Dresden, Germany}
\affil[11]{Department of Ecology, Institute of Biosciences, University of São Paulo, São Paulo, Brazil }

 \affil[*]{Corresponding author: r.martinez-garcia@hzdr.de}

 \date{}
 
\maketitle

\begin{abstract}
Wildlife-vehicle collisions (WVC) threaten both biodiversity and human safety worldwide. Despite empirical efforts to characterize the major determinants of WVC risk and optimize mitigation strategies, we still lack a theoretical framework linking traffic, landscape, and individual movement features to collision risk. Here, we introduce such a framework by leveraging recent advances in movement ecology and reaction-diffusion stochastic processes with partially absorbing boundaries. Focusing on range-resident terrestrial mammals---responsible for most fatal WVCs---we model interactions with a single linear road and derive exact expressions for key survival statistics, including mean collision time and road-induced lifespan reduction. These quantities are expressed in terms of measurable parameters, such as traffic intensity or road width, and movement parameters that can be robustly estimated from relocation data, such as home-range crossing times, home-range sizes, or distance between home-range center and road. Therefore, our work provides an effective theoretical framework integrating movement and road ecology, laying the foundation for data-driven, evidence-based strategies to mitigate WVCs and promote safer, more sustainable transportation networks.
\end{abstract}

\section{Introduction}

Collisions between wild animals and vehicles are a major source of human-wildlife conflict, threatening both biodiversity and human safety \citep{Grilo2010, Blackwell2016}. Roadkill is the second leading cause of human-induced mortality for many common and endangered vertebrate species, contributing substantially to biodiversity loss \citep{Rytwinski2015}. Wildlife-vehicle collisions (WVCs) also result in tens of thousands of human fatalities and injuries worldwide each year \citep{conover2019numbers, abra2019pay, al2013review}.  In the United States, collisions result in over 59,000 injuries and 440 fatalities annually, as well as more than 10 billion USD in economic losses \citep{conover2019numbers, huijser2008wildlife}, and the risk of any driver being involved in a WVC is roughly 1.2\%. In Germany, approximately 200,000 deer-vehicle collisions are reported annually, leading to 3,000 injured people, 50 fatalities, and causing an estimated 500 million EUR of private property damage \citep{Hothorn2012}. Recently, collisions with wild animals have also posed new challenges to the deployment of autonomous transportation systems \citep{silva_emerging_2024}. These events not only represent a direct risk to human safety, but may also disrupt animal behavior and further threaten critically endangered species \citep{da2013review, silva2021road, grilo_conservation_2021}.

Nevertheless, roads provide access to health care, education, and other services essential for economic growth and social development \citep{queiroz1992,banerjee2020}. As a result, road networks have expanded significantly in the past century and are expected to grow further in the coming decades, particularly in rural areas where WVCs may be more common \citep{meijer2018global,engert2024ghost}. Therefore, mitigating WVCs is a pressing environmental, societal, and economic challenge, critical for ensuring safe and sustainable transportation systems while safeguarding natural habitats and biodiversity \citep{gunson2015road}. Addressing this issue requires quantifying the direct impacts of roads, particularly WVCs and mortality rates, and predicting how road mortality patterns threaten the long-term viability of animal populations \citep{gunson_spatial_2011,pagany_wildlife-vehicle_2020}.

For these predictive studies to be truly effective, they must rely on an underlying theoretical framework that links mortality data to key factors such as traffic patterns, animal movement behavior, landscape features, and demographic parameters. Such theoretical frameworks are essential to identify key state variables and clarify causal relationships that cannot be disentangled from observational data alone. Purely empirical approaches often capture correlations without distinguishing among alternative mechanisms or identifying the scales at which they operate. These issues are particularly important in wildlife–vehicle collision (WVC) research. Roadkill data are affected by multiple sources of bias, including imperfect carcass detection, scavenger removal, spatial sampling biases toward accessible roads, and the lack of reliable estimates of underlying population density \citep{santos2011long, santos2016carcass,slater_assessment_2002}. Moreover, mortality counts alone do not reveal how movement behavior, space use, and traffic intensity interact to produce collision risk. Without a mechanistic link between individual movement and encounter processes, extrapolating from observed roadkill hotspots to population-level vulnerability can lead to misleading inferences \citep{ascensao_validation_2019,borda-de-agua_spatio-temporal_2011}. A theoretical framework that explicitly connects movement dynamics, landscape geometry, and traffic processes is therefore necessary to interpret empirical mortality data, correct for hidden biases, and generate robust, generalizable predictions.

The development of mechanistic theory has transformed subfields like epidemiology and community ecology, enabling quantitative predictions, extrapolation beyond observed conditions, and rigorous evaluation of management interventions \citep{Hubbell2001, Anderson1991}. Road ecology, however, remains dominated by descriptive studies with only a few efforts focusing on developing theory-driven insights \citep{Jaeger2004,borda-de-agua_spatio-temporal_2011,benard_integration_2024}. Initial attempts to study WVCs theoretically used computer simulations of animal trajectories in which simulated individuals would eventually die when crossing a road \citep{Jaeger2004}. These individual-based models addressed WVCs mechanistically, incorporating animal behaviors and vehicle movement into measurements of WVC risk \citep{benard_integration_2024}. When properly parameterized, such models can predict species' mortality risk and roadkill hotspots \citep{Hels2001,Gibbs2002,langevelde2005}. However, they rely on complex movement rules that render them mathematically intractable and limit their potential to yield explicit, generalizable relationships between movement parameters and WVC and mortality risk. For example, a recent individual-based simulation model identified correlations between road features, short-term animal behavioral responses to roads, and the expected number of WVCs over fixed time intervals \citep{benard_integration_2024}. While these results inform about how complex animal behaviors influence WVC risk, they are tied to highly specific scenarios and lack the generalizability needed to develop robust WVC risk estimators that can be informed by animal tracking data.

To overcome this problem, researchers have proposed using coarse-grained population-level models in which partial differential equations describe the spatiotemporal dynamics of total population density in the presence of roads \citep{borda-de-agua_spatio-temporal_2011}. In this approach, roads represent localized sinks that absorb the population density field at a rate that is a proxy for traffic intensity. This phenomenological description directly addresses WVCs at the population level and can provide expressions for relevant quantities, such as thresholds in road density that cause population extinctions and the corresponding expected extinction times \citep{borda-de-agua_spatio-temporal_2011}. However, these models describe movement via simple random walks, ignoring important ecological features such as range residency, which also complicates model validation using movement data \citep{Pinto2018}. This lack of a validation procedure is particularly risky when transferring these population-level models to specific systems, as they can lead to erroneous conclusions that misguide conservation efforts and WVC mitigation strategies \citep{ascensao_validation_2019}. Applied ecologists thus criticize this approach for lacking empirical validation and relying on highly unrealistic assumptions, such as animals moving infinitely fast and dying as soon as they encounter a road \citep{ascensao_validation_2019}. Despite these limitations, researchers often estimate species vulnerability based on these predictions \citep{ceiahasse_global_2017,grilo_roadkill_2020,Pinto2018}, as no better alternatives are currently available.

An improved approach to existing WVC modeling frameworks should balance a more realistic description of individual animal movement with mathematical tractability. Such an approach would offer a mechanistic understanding of how movement behavior influences key indicators of WVC risk and, by leveraging the vast amount of available animal tracking datasets, provide more quantitative predictions of species vulnerability to different road features. Here, we take a first step towards developing such a framework, focusing on range-resident species, which occupy a restricted area throughout their lifetime \citep{mueller2008search}. We study the interaction between a single animal and a linear road, treating traffic as an adjustable external parameter. Large terrestrial vertebrates, which account for most fatal WVCs and human injuries \citep{rowden2008road,sullivan2011trends,pynn2004moose,aujla2022health,conover2019numbers}, commonly exhibit this movement behavior \citep{noonan_comprehensive_2019}, making it a particularly relevant first modeling choice.


\section{An individual-level framework for wildlife-vehicle collisions}\label{sec:methods}
We introduce a general framework to quantify interactions between a moving organism and landscape linear features, which we particularize for the case of range-resident terrestrial animals interacting with vehicles traveling on a road. Wildlife-vehicle collisions are multi-step events that occur when an animal and a vehicle coincide at a road location and fail to avoid each other. Mathematically, this type of interaction can be described as a reaction-diffusion process \citep{kay_diffusion_2022,Figueiredo2025} in which reactions (collisions) occur probabilistically when a diffusion process (individual's trajectory) occupies a specific region of its phase space (the road) \citep{erban2020}. Our modeling framework thus consists of three components:\\

\noindent \textbf{Movement modeling}. We describe animal trajectories ${\bm z}(t)$ as independent realizations of a two-dimensional Ornstein-Uhlenbeck (OU) stochastic process \citep{uhlenbeck_theory_1930}. To keep the problem mathematically tractable, we assume that movement is isotropic and separable into two independent components, ${\bf z}(t)=(x(t),y(t))$. We, moreover, set the origin of coordinates at the individual home-range center, such that OU trajectories are solutions of a two-dimensional stochastic differential equation of the form
\begin{equation}\label{eq:langevin}
    \dot{{\bm z}}(t) = \bm \mu(\bm z) + \Sigma(\bm z) \, {\bm \xi}(t),
\end{equation}
where ${\bm \xi}$ is a two-dimensional Gaussian white noise with zero mean and identity covariance matrix and
\begin{equation}\label{eq:OU}
    \bm \mu(\bm z) = -\tau^{-1}\bm z; \qquad \Sigma(\bm z) = \sqrt{2\tau^{-1}}\sigma I.
\end{equation}
$I$ is the identity matrix, $\tau$ sets the characteristic time scale of the movement by fixing the home-range mean crossing time, and $\sigma^2$ sets the home-range area by modulating the intensity of the stochastic movement component \citep{fleming_fine-scale_2014}.

Equivalently, this trajectory-based description in terms of a stochastic differential equation can be formulated as an advection–diffusion (Fokker–Planck) equation for the probability distribution of the animal's location. In this representation, the diffusion coefficient is $\sigma^2/\tau$ and the advection velocity $(\bm{z}-\bm{\mu})/\tau$ \citep{Menezes2025, smouse2010stochastic}. This description is closer to a mechanistic home-range analysis framework \citep{moorcroft2013mechanistic}.
\\

\begin{figure}
\centering
\includegraphics[width={0.7\linewidth}]{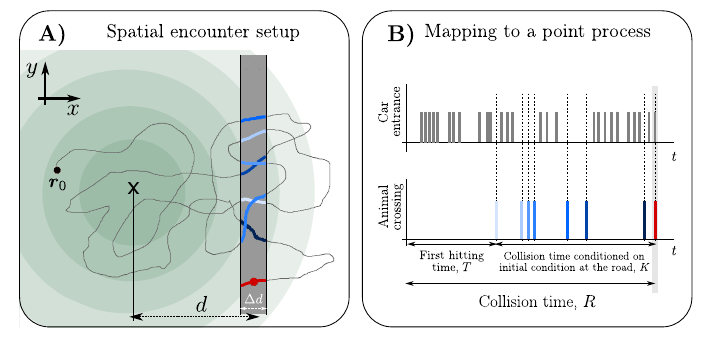}
\caption{A) Schematic representation of the wildlife-vehicle interaction problem. The green shaded area represents the animal space utilization function, with darker green accounting for regions with a higher occupancy probability. A linear road of width $\Delta d$, where cars appear following a Poisson point process with rate $\nu_0$, crosses the animal home range at distance $d$ from its center. We are interested in computing the collision times between these cars and animal trajectories starting from a random initial condition, $\bm{r}_0$, sampled from the animal utilization function. (B) This encounter process can be mapped to a time process in which traffic intensity is defined by a local Poisson process with rate $\nu_0$ on the road, and the features of animal trajectories in space define the time statistics of its road crossing events. Within this framework, the animal may cross the road several times (blue lines in panels A and B) before colliding with a car (red line).} \label{fig:mapping}
\end{figure}

\noindent \textbf{Road and traffic modeling.} We consider that a linear road of width $\Delta d$ crosses the landscape at a distance $d$ from the home-range center. Without loss of generality and to keep mathematical simplicity, we define the $y$ axis parallel to the road (Fig.\,\ref{fig:mapping}A). Mathematically, such a road is defined as the region of the landscape
\begin{equation}\label{eq:roaddef}
\Omega = [d - \Delta d/2, d + \Delta d/2]\times \mathbb R \subset \mathbb R^2.
\end{equation}
We \textcolor{blue}{focus primarily on}  animals whose home ranges are such that the typical length scale of motion is much larger than the road width, $\sigma\gg\Delta d$, which is true for most mid- to large-sized mammals 
\citep{noonan_comprehensive_2019}.For completeness, we treat the general case in which this separation of scales does not hold in Appendix \ref{app:roadwidth}.

For traffic dynamics, we assume that the movement of vehicles is approximately mutually independent and much faster than animal movement, which allows us not to resolve traffic in space. Instead, we model road traffic as a sequence of discrete temporal events, each representing the appearance of a vehicle. Because we assume that these discrete events are mutually independent, we can represent traffic volume in terms of a homogeneous Poisson point process of incoming vehicles at a rate $\nu_0$.\\

\noindent \textbf{Collision modeling.} In real-world scenarios, WVCs are multi-step complex processes resulting from traffic volume, and how animals and drivers respond to each other's presence \citep{jacobson2016behavior,pagany_wildlife-vehicle_2020}. Given a proportion $q$ of accidents \textit{per} incoming vehicle and a rate of incoming vehicles $\nu_0$, potential collisions occur at a rate $\nu = q\nu_0$. This rate defines the time sequence of potential collisions as a local Poisson point process on the road $C_\Omega = \left\{c_i\right\}$ $i\in\mathbb N$. Because $C_\Omega$ has homogeneous and stationary increments, it defines a sequence of times of possible WVCs that are only realized if the animal is at the road---i.e., occupies $\Omega$---at time $c_i$ (Fig.\,\ref{fig:mapping}B). Considering this definition for $C_\Omega$, the collision time $R$ is the stopping time for the movement process defined by the first coincidence of a potential collision event with the occupation of the region $\Omega$,
\begin{equation}
\label{eq:collisiontime}
R = \min \left\{t \in [0, \infty) \;|\; t \in C_{\Omega} \text{ and } \boldsymbol z(t) \in \Omega  \right\}.
\end{equation}

\noindent The general setting for the study of this collision time is that of diffusion in the presence of reactive domains or semi-permeable barriers  \citep{kenkre_theory_2014,kay_diffusion_2022,Figueiredo2025}. Here, we study how the characteristic time scale of the animal movement and the collision rate within the road jointly determine the collision time. For sufficiently small $\Omega$, the strong Markov property allows the approximate decomposition of the collision time $R$ into two independent terms,
\begin{equation}\label{eq:decomposition}
    R = T + K,
\end{equation}
\noindent where $T$ is the animal's first hitting-time to the road, $\Omega$, given the initial conditions. That is, 
\begin{equation}
    T = \min \{t \in [0,\infty) | \bm z(t) \in \Omega\},
\end{equation}
which does not depend at all on the properties of the reactive domain. On the other hand, $K$ is the collision time given the initial condition that the animal is at the road $\Omega$ (Fig.\,\ref{fig:mapping}B).

To obtain exact expressions for some of the statistical properties of the collision times, we reduce this two-dimensional modeling setup to a one-dimensional problem by exploiting some of the assumptions we made regarding landscape geometry and movement behavior. In particular, we model the road as a straight line aligned with one of the coordinate axes and consider animal movement components along each of these axes independent. Additionally, we do not resolve car movement within the road and consider that the road is either occupied by a car or empty following a Poisson process. Under these assumptions, displacements in the $y$ coordinate (parallel to the road; Fig.\,\ref{fig:mapping}) are irrelevant for the road crossing events and we can obtain statistics of the collision times by working with the probability density $P(x,t)$ in the direction perpendicular to the road. Specifically, we obtain exact expressions for the collision time distribution and use this result to compute two metrics of WVC risk. First, the expected collision time, which defines a characteristic time scale at which the animal will eventually die on the road. Second, the expected animal lifetime reduction due to mortality on the road, which is based on the extent to which death by collision increases the animal's chances of dying above its background death rate.

\section{Results}


\subsection{The wildlife-vehicle collision time distribution}\label{sec:distribution}


The time-homogeneous Poisson process we use to model potential collision times remains a Poisson process with the same rate when restricted to the intervals during which the animal occupies the road segment $\Omega$. If we assume that the animal’s trajectory terminates upon collision and enters a cemetery or absorbing state, then the transition probability $P(x,t|x_0)$, conditioned on a deterministic initial condition $P(x, 0) = \delta(x-x_0)$, satisfies a backwards (or adjoint) Fokker-Planck equation in $x_0$ \citep{gardiner_stochastic_2009}
\begin{align} \label{eq:FPEback}
\frac{\partial P(x, t|x_0)}{\partial t} &= \left[ \mu(x_0)\frac{\partial}{\partial x_0} + \frac{g(x_0)}{2}\frac{\partial^2}{\partial x_0^2} - \omega(x_0)\right]P(x,t|x_0) \nonumber \\
&= \left[\hat L^\dagger_{x_0} - \omega(x_0)\right]P(x,t|x_0) 
\end{align}
where $\hat L^\dagger_{x_0}$ is an adjoint Fokker-Planck operator and $\mu(x) = \mu_1(\bm z)$, $g(x) = (\Sigma^\intercal\Sigma)_{11}(\bm z)/2$. 
In this adjoint equation, the variable $x$ is only a parameter and it can be integrated over. This operation defines the survival function of the stochastic trajectory, $S(t|x_0)$ \citep{gardiner_stochastic_2009}. This survival function  quantifies the probability that the trajectory describing animal movement is still \textit{alive}, and thus measures the probability that a WVC has not taken place at time $t$ given an initial animal location $x_0$. Mathematically, this survival function is the complementary cumulative distribution function of $R$,
\begin{equation}
    S(t|x_0) = \int P(x,t|x_0)\text{d}x,
\end{equation}
and satisfies the same backward equation as $P(x,t|x_0)$, with initial condition $S(0|x_0) = 1$ and long-time convergence to zero $S(t \rightarrow \infty|x_0) = 0$ due to recurrence of the one-dimensional OU. In ecological terms, this recurrence implies that the animal will always collide with a vehicle if observed for long enough. 

Finally, the assumption that the scale of motion is much larger than the road width allows us to approximate our original description of the road in Eq.\,\eqref{eq:roaddef}, i.e. $\omega(x)$ uniform in the interval $[d - \Delta d/2, d + \Delta d/2]$, by a point sink
\begin{equation}\label{eq:deltaapprox}
    \omega(x) = \nu\mathbbm 1 [d - \Delta d/2 < x < d + \Delta d/2] \approx \eta \delta(d-x),
\end{equation}
where we have introduced an effective traffic intensity $\eta=\nu \Delta d$ that accounts for both road width and the rate at which cars enter it and has dimensions of velocity \citep{kay_defect_2022,kenkre_montroll_2021,Figueiredo2025}. As we discuss in more detail in Appendix \ref{sec:micro}, the infinitely narrow road limit, $\Delta d\rightarrow 0$, must be treated carefully in numerical simulations because OU trajectories are not differentiable. However, following the literature \citep{erban_reactive_2007}, we can relate the death probability per crossing, a typical output of individual-based computational models of WVCs \citep{Hels2001,Gibbs2002}, to the effective traffic intensity $\eta$
\begin{equation}
    P_\times = \sqrt{\frac{\pi \tau \Delta t}{4\sigma^2}}\eta,
\end{equation}
where $\Delta t$ is the time discretization used to simulate the animal trajectory.

In the limit where the road is a point sink, the problem of determining the statistics of $R$ can thus be reduced to analyzing the equation
\begin{equation}\label{eq:survival}
    \frac{\partial S(t|x_0)}{\partial t} = \left[\hat L^\dagger_{x_0} - \eta \delta(d - x)\right]S(t|x_0),
\end{equation}
(see Appendix\,\ref{app:roadwidth} for a treatment of the full problem with finite road width $\Delta d$). This equation can be equivalently written in Laplace space as an equation for the moment generating function (MGF) of R, $\langle e^{-sR}\rangle_{x_0}$, which is the Laplace transform of the probability density function $\phi(t|x_0) = -\partial S/{\partial t}$. In Laplace space, Eq.\,\eqref{eq:survival} becomes
\begin{equation}\label{eq:MGF}
    \left[\hat L^\dagger_{x_0} - s - \eta \delta(d-x_0)\right]\langle e^{-sR}\rangle_{x_0} = -\eta \delta(d - x_0),
\end{equation}
where $s$ is the Laplace-space frequency variable, and we henceforth denote by $\langle \cdot \rangle_{x_0}$ the expectation conditioned on a deterministic initial condition $x_0$ and drop the subscript when such an initial condition is sampled from the stationary probability distribution of the one-dimensional OU process, $p_{\mathrm{st}}(x)$. In the limit of infinite effective traffic intensity $\eta\rightarrow\infty$, the probability sink accounting for road mortality in Eq.\,\eqref{eq:MGF} can be replaced by Dirichlet boundary conditions and \eqref{eq:MGF} becomes an equation for the MGF of the first-hitting time (i.e., the time at which the animal trajectory reaches the road for the first time), $\langle {\rm e}^{-sT}\rangle_{x_0}$. Expanding this equation in series of $s$, we can thus obtain the mean first hitting time, $\langle T\rangle$. 

\begin{figure}
\centering
\includegraphics[width={0.75\linewidth}]{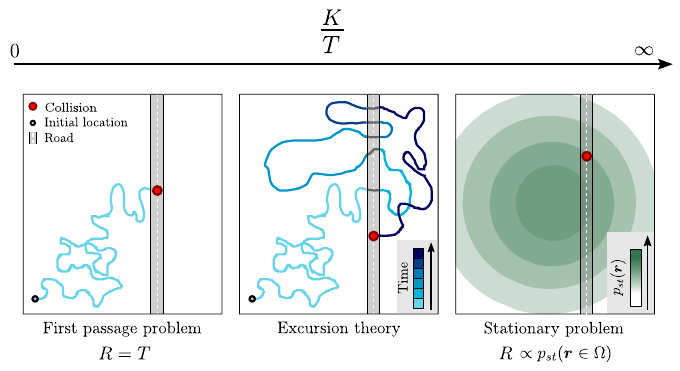}
\caption{Scheme of the decomposition of the interaction time. Depending on the ratio between the first-hitting time to the road, $T$ and the collision time conditioned on initial condition at the road $K$, we distinguish three regimes. For $T\ll K$ (left panel), the encounter reduces to a first-passage problem; for $T\gg K$ (right panel), the collision time is fully dominated by dynamics close to the road and thus proportional to the road occupation probability. Between these two limits (central panel),the encounter process can be studied using excursion theory}\label{fig:summary}
\end{figure}

For a finite $\eta$, following the steps in \citep{Figueiredo2025} and references therein, we can solve Eq.\,\eqref{eq:MGF} to obtain the moment generating function of the collision time $R$,
\begin{align}\label{eq:MGFstationary}
\langle e^{-s R} \rangle_{x_0} &= \frac{\langle e^{-sT}\rangle_{x_0}}{1 + \frac{s}{\omega_d}\langle e^{-sT}\rangle},
\end{align}
where we have defined $\omega_d\equiv \eta \, p_{\mathrm{st}}(d)$. If we consider that the stochastic process modeling animal movement has a characteristic time scale $T$, this expression for the MGF of $R$ formalizes the decomposition of the collision time we introduced in Eq.\,\eqref{eq:decomposition} by considering two limits. First, for $\langle T \rangle \gg \omega_d^{-1}$, the denominator in Eq.\,\eqref{eq:MGFstationary} reduces to one and $R\approx T$, which means that the collision-time statistics is fully given by the first hitting time distribution. On the opposite end, if $\langle T \rangle \ll \omega_d^{-1}$, then Eq.\,\eqref{eq:MGFstationary} converges to the MGF of an exponential distribution and the collision time becomes an exponentially distributed random variable $R\sim \mathrm{Exp}(\omega_d)$ with $\omega_d = \eta \, p_{\mathrm{st}}(d)$. Therefore, in this limit, the collision times only depend on the properties of the traffic, the road, and the animal range distribution evaluated at the road location.
In this limit, WVCs are given by a stationary process in which collisions are uniformly distributed in time with a constant rate given by a combination of the traffic parameter and road occupation probability (Fig.\,\ref{fig:summary}). Additionally, expanding Eq.\,\eqref{eq:MGFstationary} in $s$, we find that a first correction to the stationary ergodic regime is still given by an exponential variable whose rate is now the harmonic mean of the stationary reaction rate and the inverse of the mean first passage time,
\begin{equation}\label{eq:exponentialapprox}
    T \approxsim \text{Exp}\left(\frac{1}{\omega_d^{-1} + \langle T \rangle}\right).
\end{equation}
Beyond this first order correction, the collision time can be more generally obtained using excursion theory \citep{ito_poisson_1969, ito_poisson_1971}. 

\subsection{Expected wildlife-vehicle collision time} 

Starting from the decomposition in Eq.\,\eqref{eq:decomposition}, we can calculate the mean collision time, conditioned on initial condition sampled from the one-dimensional OU stationary probability density function $p_{st}(x)$, as $\langle R\rangle = \langle T \rangle + \langle K \rangle$. In this section, we obtain analytical expressions for both contributions to the mean collision time and analyze how the different movement and traffic parameters determine this WVC risk indicator. 
The first-hitting time distribution of the OU process conditioned on a deterministic initialization can be found by solving the parabolic cylinder equation that results from taking the $\eta\rightarrow\infty$ limit in \eqref{eq:MGF}. The properties and representations of this first-hitting time distribution have been explored in detail \citep{siegert_first_1951,sato_moments_1978,ricciardi_first-passage-time_1988,alili_representations_2005,lipton_first_2018,hartich_interlacing_2019}, but we provide a full derivation in Appendix \ref{app:smfpt}. Written in terms of the parabolic cylinder function $D_{-\nu}(x)$, the MGF of the first hitting time is
\begin{align}\label{eq:mgfstationaryOU}
    \langle e^{-s T} \rangle = \frac{e^{-\frac{d^2}{2\sigma^2}}}{\Gamma(s \tau + 1)D_{-s\tau}\left(-\frac{d}{\sigma}\right)D_{-s\tau}\left(\frac{d}{\sigma}\right)},
\end{align}
which one can expand in series of $s$ to explicitly evaluate the first moment and thus obtain the stationary mean first hitting time (see Appendix \ref{app:smfpt} for the details of the calculation),
\begin{equation}\label{eq:stationaryMFPTOU}
    \langle T \rangle = \tau\log 2 + \tau\frac{d^2}{\sigma^2} \hypgeo{2}{2}\left(1,1;\frac{3}{2},2;\frac{d^2}{2\sigma^2}\right),
\end{equation}
where $\hypgeo{2}{2}$ is a generalized hypergeometric function that is equal to one when $d/\sigma=0$ and increases as $d/\sigma\rightarrow\infty$. The second contribution to the collision time, the collision time conditioned on initialization at the road $K$, can be obtained by simply evaluating $\omega_d^{-1} = [\eta \, p_{\mathrm{st}}(d)]^{-1}$ (see Appendix \,\ref{sec:momentK} and \citep{Figueiredo2025} for a full derivation). Considering the stationary probability density function of the OU movement model, we obtain \begin{equation}\label{eq:OUstat}
    \langle K \rangle = \omega_d^{-1} = \frac{\sqrt{2\pi \sigma^2}}{\eta}\mathrm{e}^{\frac{d^2}{2 \sigma^2}}.
\end{equation}

\begin{figure}[ht]
\centering
\includegraphics[width={0.7\linewidth}]{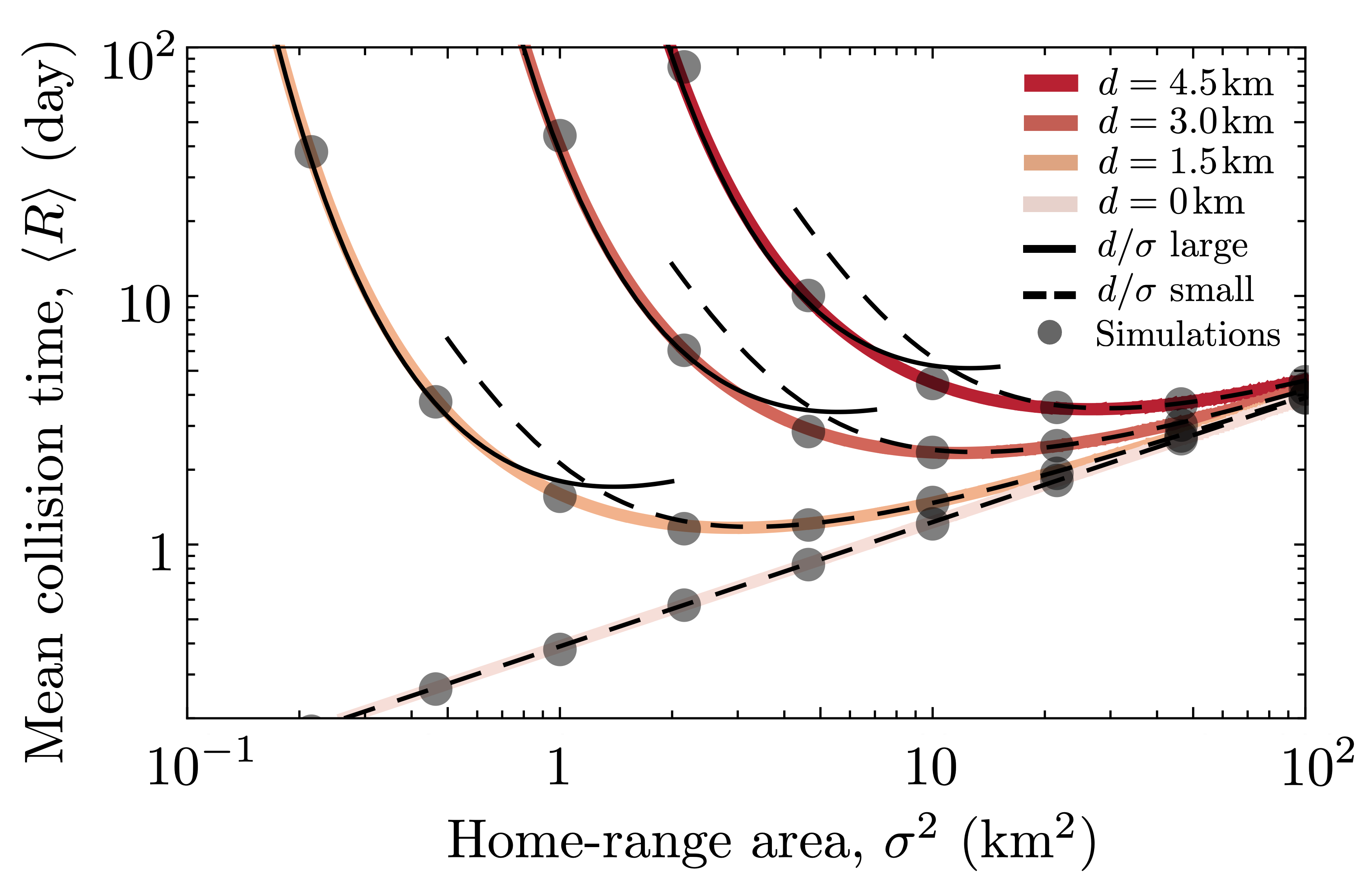}
\caption{\label{fig:val} (Log-log scale). Validation of theoretical results for the mean collision time, $\langle R \rangle$, as a function of the home-range area $\sigma^2$ and varying the distance between the road and the home-range center. Solid color lines show the exact result, while solid-black and dashed-black lines show, respectively, the large-$d/\sigma$ and small-$d/\sigma$ expansions. Gray dots are obtained from direct numerical simulations of the reaction-diffusion stochastic dynamics (see Appendix\,\ref{sec:micro} for details on the implementation). Parameters: $\eta = 10\,$km/day, $\tau$ varies such that $\sigma/\tau=5\,$km/day.}
\end{figure}

Because the generalized hypergeometric function is always positive, the first immediate result from Eqs.\,\eqref{eq:stationaryMFPTOU}-\eqref{eq:OUstat} is that, everything else held constant, the mean collision time increases with increasing home-range crossing time and increasing the distance between the road and the home-range center. Additionally, to validate our theoretical results, we compared the mean collision time $\langle R \rangle$ obtained from Eqs.\,\eqref{eq:stationaryMFPTOU} and \eqref{eq:OUstat} with the values obtained from numerical simulations of the individual-level process (see Appendix\,\ref{sec:micro} for details on the numerical simulations). We performed this comparison for various distances between the road and the home-range center, $d$, and home-range size, $\sigma$, while keeping the effective traffic intensity $\eta$ and characteristic animal velocity, $v\sim\sigma/\tau$, constant. For low values of $d$, larger home ranges lead to higher collision times, while larger values of $d$ revert this trend (Fig.\,\ref{fig:val}).

Two particularly relevant limits, representative of how different species use roads, are those in which the road is sufficiently far from the home-range center and hence rarely visited by the animal, and the limit in which home ranges are established close to roads \citep{Seigleferrand2022}. We can approximate the behavior of the generalized hypergeometric function in these two limits to obtain simpler expressions for the mean collision time. If we consider the limit in which roads are sufficiently far from the animal home-range center, $d\geq 2\sigma$, we can use the asymptote of the hypergeometric function \citep{volkmer_note_2013} and approximate the mean first hitting time by,
\begin{equation}\label{eq:appstationaryMFPTOU}
    \langle T \rangle \sim \tau \frac{\sqrt{2\pi\sigma^2}}{d}e^{\frac{d^2}{2\sigma^2}}.
\end{equation}
$\langle R\rangle$ can thus be approximated by \eqref{eq:appstationaryMFPTOU} and \eqref{eq:OUstat} (solid black lines in Fig.\,\ref{fig:val}). Alternatively, when home ranges are placed close to the road and $d\ll \sigma$, the mean first hitting time at lowest order in $d/\sigma$ is
\begin{equation}\label{eq:appstationaryshortMFPTOU}
    \langle T \rangle = \tau\log 2 + \tau \frac{d^2}{\sigma^2},
\end{equation}
which also provides a good approximation to the mean collision time in this limit (dashed black lines Fig.\,\ref{fig:val}).

\begin{figure}[ht]
\centering
\includegraphics[width={\linewidth}]{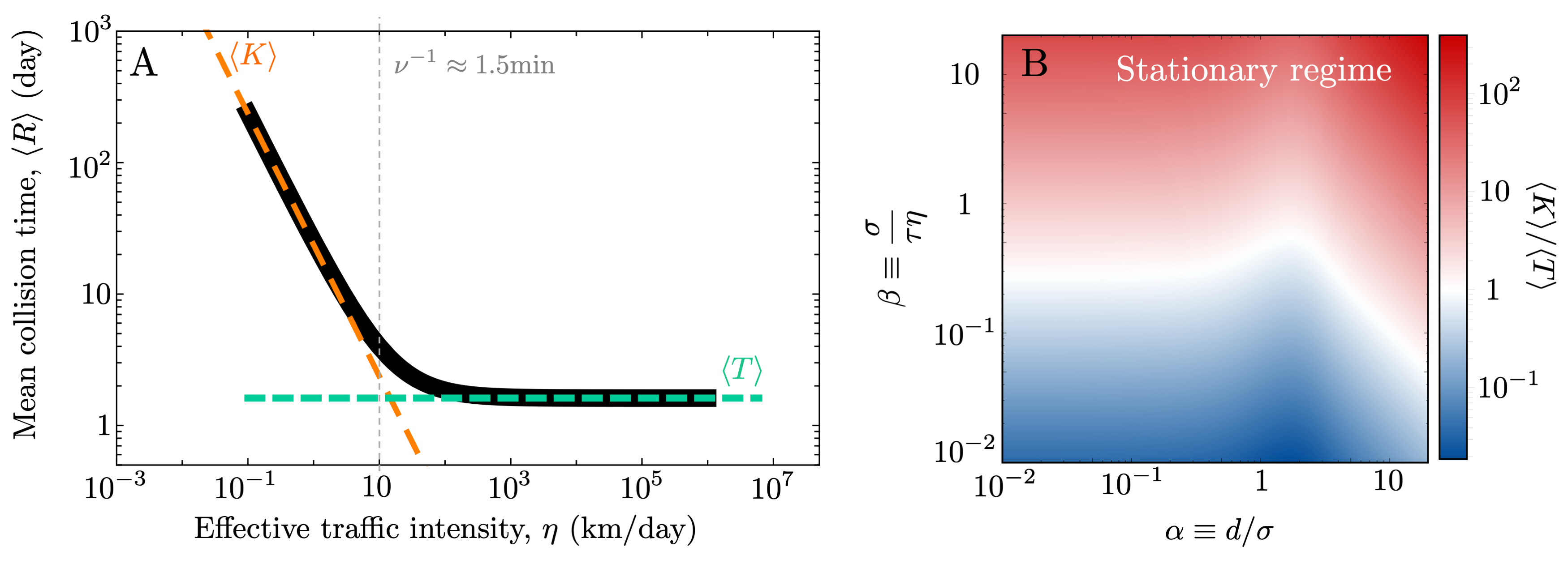}
\caption{\label{fig:lim} A) Regime separation in the mean collision time $\langle R \rangle$ as a function of the effective traffic intensity $\eta$. For a parameter set $\sigma^2=100\,\mathrm{km}^2$ and $\tau=2\,\mathrm{day}$, animal-vehicle collisions can be studied using the animal range distribution when the average time between car appearances is lower than $15\,\mathrm{min}$, assuming a road width of $10\,\mathrm{m}$. B) $\langle K \rangle / \langle T \rangle$ as a function of the dimensionless parameters $\alpha\equiv d/\sigma$ and $\beta \equiv \sigma/(\eta \tau)$. When this quantity is large (red region), animal-vehicle collisions can be studied in terms of animal range distributions.}
\end{figure}

Finally, we use the expressions for $\langle T \rangle$ and $\langle K \rangle$, \eqref{eq:stationaryMFPTOU} and \eqref{eq:OUstat} respectively,  to explore how the diffusion and reaction-limited regimes discussed in Section \ref{sec:distribution} are defined by traffic intensity and other measurable parameters. First, to build a quantitative intuition on how these two regimes are settled within a realistic range of parameters, we consider a hypothetical animal with a range distribution area $\sigma^2=100\,\mathrm{km}^2$ and average home-range crossing time $\tau=2\,\mathrm{day}$ whose home-range is crossed by a road of width $10\,\mathrm{m}$ (so the $\Delta d\ll \sigma$ condition is met) at a distance $d=4\,\mathrm{km}$ from the home-range center. For this parameterization, we compute the average collision time as a function of the effective traffic intensity $\eta$, or equivalently---because we hold $\Delta d$ constant---the intensity of the Poisson process that models traffic $\nu$ (Fig.\,\ref{fig:lim}A). At high traffic intensity (large $\eta$), the collision time is controlled by the mean first hitting time, $\langle T \rangle$. However, when the traffic rate is low enough---the average time between two consecutive car entrances in the road is approximately $1.5\,\mathrm{min}$---the average collision time is dominated by $\langle K \rangle$ and the interaction between the animal and the road can be studied as a stationary problem. To define this transition more generally, we introduce two dimensionless parameters
\begin{equation}\label{eq:defalphabeta}
\alpha \equiv \frac{d}{\sigma}, \qquad  \beta \equiv \frac{\sigma}{\tau \eta}.
\end{equation}
Using these parameters, we write a scaled collision time as
\begin{equation}
    \frac{\langle R\rangle}{\tau} = \log 2 + \alpha^2 \hypgeo{2}{2}\left(1,1;\frac{3}{2},2;\frac{\alpha^2}{2}\right) + \sqrt{2 \pi}\beta\mathrm{e}^{\frac{\alpha^2}{2}},
\end{equation}
where the first two contributions on the right side correspond to $\langle T\rangle/\tau$ and the last one to $\langle K\rangle/\tau$. The ratio $\langle T\rangle/\langle K\rangle$ delimits the range of movement ($\sigma, \tau$), landscape ($d, \Delta d$), and traffic parameters ($\nu$) for which WVCs can be studied as a stationary problem and for which we must consider the first hitting time. For small $\alpha$, $\langle K \rangle/\langle T \rangle$ does not depend on $\alpha$ and WVCs are a stationary problem when $\beta\gtrapprox 1$ or, equivalently, $\sigma/\tau \gtrapprox \eta$. For large $\alpha$, using Eq.\,\eqref{eq:appstationaryMFPTOU} to approximate $\langle T \rangle$, we get $\langle K \rangle/\langle T \rangle \sim \alpha \beta \gg 1$, which allows us to evaluate whether the stationary approximation is accurate for a given set of parameters. Using the definition of $\alpha$ and $\beta$, we obtain in the large-$\alpha$ limit $\langle K \rangle/\langle T \rangle \sim d/(\tau\eta)$, which does not depend on the animal home-range size.

\subsection{Measuring road mortality in terms of expected lifetime reduction}

 The presence of a road in the animal habitat introduces an additional hazard on top of the usual processes that determine animal mortality. We next quantify how such an additional hazard impacts the animal's lifetime by computing the expected lifetime reduction (i.e., how likely the animal is to die in the road before it does due to any other cause). Throughout these calculations, we maintain the assumption that the road width is negligible relative to the spatial scales of animal movement, which yields accurate estimates of the probability of premature death except in scenarios where intrinsic death rates are very large or roads are very far from the animal's home-range center. In these limits, the impact of the road on an animal's lifetime is already very small and thus provide less interesting scenarios (see Appendix \ref{app:roadwidth} for this analysis).
 
 Assuming all these preexisting hazards lead to an exponential survival curve, the basal individual lifetime is an exponentially distributed random variable, $\Delta \sim \text{Exp}(\delta)$. Because $\Delta$ and $R$ are independent random variables, we obtain the probability of animal \textit{premature} death on the road using the joint probability distribution of $R$ and $\Delta$,
\begin{equation}\label{eq:delta-t0}
    \mathbb P[\Delta > R] = \int_{0}^{\infty} \text dt \int_{0}^{t} \text dt' \delta \mathrm{e}^{-\delta t'} \phi(t).
\end{equation}
Performing the integral on $t'$ in Eq.\,\eqref{eq:delta-t0}, we get
\begin{equation}\label{eq:delta-t}
 \mathbb P[\Delta > R] = \int_0^\infty e^{-\delta t}\phi(t)\text dt = \langle e^{-\delta R} \rangle,
 \end{equation}
that gives the MGF of $R$ a natural probabilistic interpretation in terms of the probability of premature death due to the presence of the spatial hazard. The actual lifetime under this interpretation is the minimum of the intrinsic lifetime and the collision time,
\begin{equation}
    T_{\Delta} \equiv \min(\Delta,R),
\end{equation}
so that its complementary cumulative distribution function or survival function, again because $\Delta$ and $R$ are independent, is simply the product of the survival probability associated with either process
\begin{equation}
    S_{T_\Delta}(t) = S_R(t)e^{-\delta t},
\end{equation}
where we have already considered that the survival function associated with $\Delta$ is exponential with rate $\delta$. Expressed in these terms, the reduction in average lifespan due to the road is identical to the probability of premature death in Eq.\,\eqref{eq:delta-t},
\begin{equation}
     \frac{\langle \Delta - T_\Delta \rangle}{\langle \Delta \rangle} = \langle e^{-\delta R}\rangle.
\end{equation}

\begin{figure}[ht]
\centering
\includegraphics[width={0.8\linewidth}]{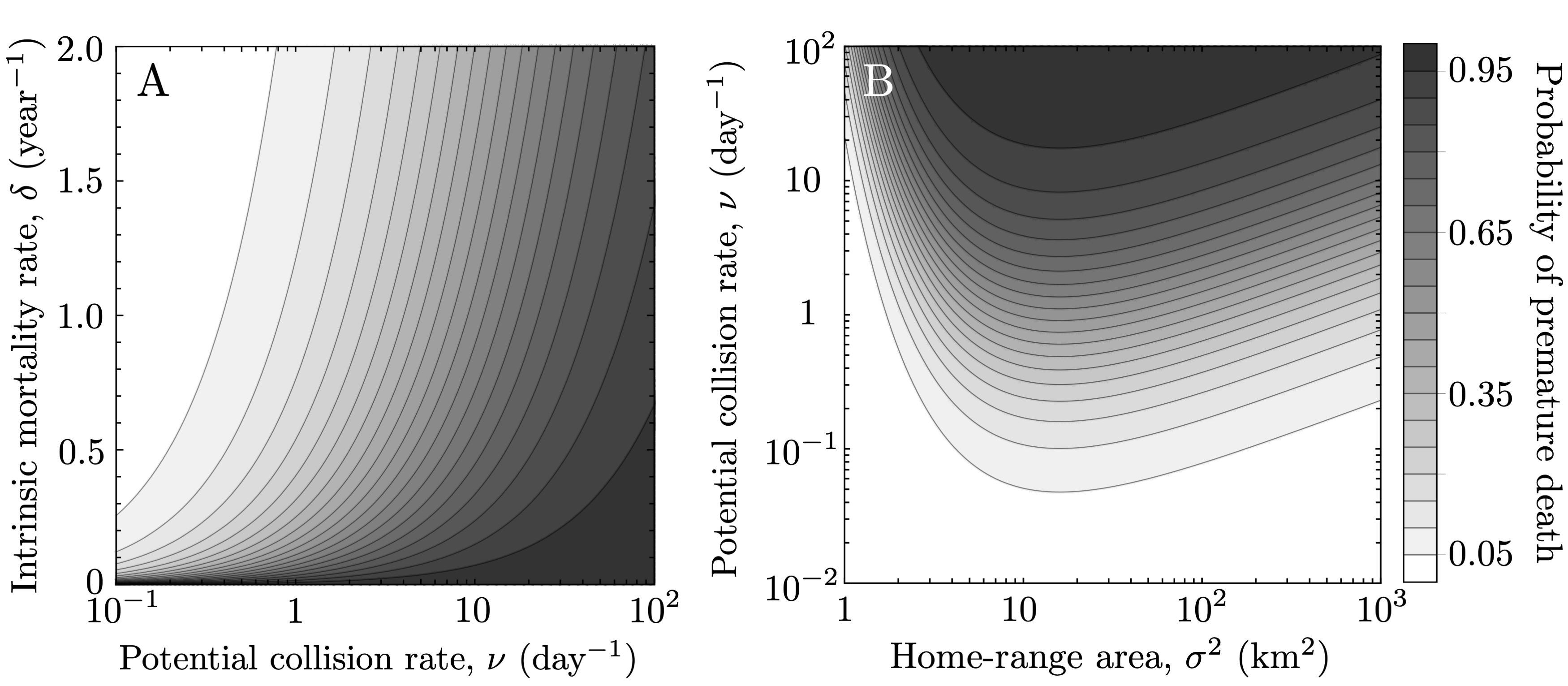}
\caption{\label{fig:prem}A) (Log-linear scale) Probability of premature death, or equivalently, reduction in average lifespan, as a function of the intensity of the hazards that can cause animal death: effective traffic intensity $\eta$ and intrinsic death rate, $\delta$. Parameters like in Fig.\,\ref{fig:lim}A: $\sigma^2=100\,\mathrm{km}^2$, $\tau=2\,\mathrm{day}$, $\Delta d = 10\,\mathrm{m}$, $d=4\,\mathrm{km}$. B) (Log-log scale)
Probability of premature death as a function of potential collision rate and home-range area. Parameters: $\delta = 0.2\, \mathrm{year}^{-1}$, $d=4\,\mathrm{km}$, $\Delta d = 10\,\mathrm{m}$, and $\tau$ varies with $\sigma$ to keep a characteristic animal velocity $\sigma/\tau= 5\,\mathrm{km/day}$ constant.}
\end{figure}

In the majority of realistic scenarios, the intrinsic mortality rate will be of the order of years$^{-1}$ while $T$ will typically be of the order of days. Therefore, because $\delta \ll \langle T \rangle^{-1}$, we can use Eq.\,\eqref{eq:exponentialapprox} to approximate the MGF of $R$ and write the reduction in average lifespan as
\begin{equation}\label{eq:reduction}
     \frac{\langle \Delta - T_\Delta \rangle}{\langle \Delta \rangle} = \langle e^{-\delta R}\rangle \approx \frac{1}{1 + \delta\left(\omega_d^{-1} + \langle T \rangle\right)},
\end{equation}
which we can evaluate simply by replacing the expressions for $\omega_d^{-1}$ and the mean first hitting time obtained in Eqs.\,\eqref{eq:stationaryMFPTOU} and \eqref{eq:OUstat}, respectively. As expected, the probability of premature death due to the road increases with increasing effective traffic intensity and decreasing intrinsic mortality rate (Fig.\,\ref{fig:prem}A). Additionally, we also looked at how the probability of premature death changes with the size of the home range and the intensity of traffic for a hypothetical individual whose characteristic velocity $\sigma/\tau$ and the intrinsic mortality rate we maintained constant. Because the mean collision time has a minimum at intermediate home-range sizes (darker red lines in Fig.\,\ref{fig:val}), the probability of premature death also maximizes at intermediate home-range sizes (Fig.\,\ref{fig:prem}B). This result suggests that a complex interaction between range-residency parameters and landscape features determines animal survival.

Finally, if we introduce the dimensionless parameters $\alpha$ and $\beta$ defined in \eqref{eq:defalphabeta} and additionally consider the limit of distant roads, $d \geq 2\sigma$ ($\alpha\geq 2)$, we can use the asymptote to the hypergeometric function in Eq.\,\eqref{eq:appstationaryMFPTOU} to approximate the mean first-hitting time and the probability that roadkill occurs before natural death is
\begin{equation}
    \frac{\langle \Delta - T_\Delta \rangle}{\langle \Delta \rangle} \approx \frac{1}{1 + \sqrt{2\pi}\gamma \left(\beta + \alpha^{-1} \right)e^{\frac{\alpha^2}{2}}},
\end{equation}
where we have defined a third dimensionless parameter, $\gamma\equiv \tau \delta$, accounting for the ratio between the home-range crossing and intrinsic mortality time scales. Similarly, we can consider the limit where home ranges are established close to roads and hence $\alpha$ is small. In this limit, retaining the lowest order terms in $\alpha$ in the Taylor expansion of Eq.\,\eqref{eq:reduction}, we obtain
\begin{equation}
    \frac{\langle \Delta - T_\Delta \rangle}{\langle \Delta \rangle} \approx \frac{1}{ 1 + \gamma \big[2\pi\beta + \log 2 + (1+2\pi\beta)\alpha^2\big]}.
\end{equation}

\section{Discussion}
Wildlife-vehicle collisions are multi-step processes in which an animal and a vehicle must simultaneously occupy a road location and fail to avoid each other. Mathematically, these processes can be studied as a reaction-diffusion process, in which reactions (collisions) occur at a given rate, provided that a diffusion process (individual's movement) occupies a specific region of its phase space (the road) \citep{erban2020}. Building on recent advances in theoretical studies of animal encounters through reaction-diffusion processes \citep{gurarie2013, martinez-garcia_how_2020, das_misconceptions_2023, Figueiredo2025} and leveraging well-established techniques from condensed matter physics \citep{Montroll1955,kenkre_montroll_2021,kay_defect_2022}, we developed the first mathematically tractable framework to study animal-vehicle collisions at the level of individual movement and, to further demonstrate how this new framework captures key features of WVC dynamics, we compared our theoretical results with numerical simulations of the collision process.

We focused on range-resident terrestrial animals, which account for the highest number of human fatalities and injuries caused by WVCs \citep{rowden2008road,aujla2022health,conover2019numbers}, and derived expressions for the expected collision time and the probability that a collision occurs before the animal dies from other causes. In our framework, this probability of \textit{premature} death translates to a reduction in the individual's average lifespan due to road mortality. The expected collision time depends in non-trivial ways on easily measurable parameters, such as traffic volume or road width, as well as movement parameters for which robust statistical estimators exist, such as home-range size and home-range crossing rates \citep{calabrese2016}. We further derived relationships between the effective traffic intensity, $\eta$, and the probability of death per crossing, a typical output of fine-scale simulation models \citep{Hels2001,benard_integration_2024}. The reduction in average lifespan requires knowledge of a species' or population's lifespan, which can be estimated from survivorship curves \citep{caughley1966,lynch2009}. Therefore, our results provide a foundation for deriving statistical estimators that quantify road-induced animal mortality patterns based on traffic volume parameters and animal movement behavior. However, such an effort would entail substantial additional technical work and is thus well beyond the scope of this paper.

Beyond its mathematical tractability, our framework shows that traffic intensity defines a shift between two qualitatively different mortality regimes, each implying fundamentally different management strategies depending on the ratio between movement time scales and traffic intensity. When traffic intensity is high, WVCs occur on a time scale much shorter than the animal's characteristic movement time. In this diffusion-limited regime \citep{Figueiredo2025}, survival is determined almost entirely by when animals first hit roads, which act as nearly absorbing boundaries. In this regime, fine-scale behaviors at the road (e.g., crossing speed, vigilance, or time spent foraging on verges) have little influence on long-term survival, and mitigation efforts should primarily aim to reduce encounter probability through spatial planning, fencing, or rerouting. Estimating collision times in these scenarios is very challenging because it requires the estimation of first-hitting times from statistics of animal trajectories.

In contrast, at low traffic intensity, mortality does not depend on how often animals reach roads, but on how long they spend on them. In this reaction-limited regime \citep{Figueiredo2025}, behavioral avoidance, reduced road-use time, or temporal shifts in activity patterns may substantially reduce mortality risk, because fine-scale trajectory details become irrelevant and collision risk depends only on the stationary space-use distribution.

This property of the collision time simplifies its statistical estimation by extending existing tools for estimating animal encounters through home-range overlap \citep{noonan_estimating_2021,fagan2024}. An additional notable feature of this regime is that the reduction in average lifespan, as a function of increasing distance to the road, decays proportionally to the tail of the range distribution itself [see Eq.\,\eqref{eq:MGFstationary}]. In a risk mitigation context, this result suggests that changes in the overlap between a road and an animal's range may have a strongly non-linear effect on excess mortality. Additionally, increasing home-range size while keeping the distance from the home-range center to the road constant does not necessarily increase mortality, because animals use space within their home ranges unevenly, and mortality depends on the overlap between individual range distributions and the road. This highlights the importance of accurately quantifying space use as a distribution.

The existence of this regime in which individual mortality on the road can be described in terms of a Poisson process also facilitates upscaling our framework to larger population sizes while retaining inter-individual variability in movement behavior \citep{Menezes2025}. Although this upscaling presents mathematical challenges, it is essential for developing spatially explicit population dynamics that account for individual differences in space use and their interactions with roads and other human infrastructures. This new class of models would provide a foundation for refining current estimators of species vulnerability to roads and for evaluating the broader ecological impacts of roadkill on biodiversity \citep{ceiahasse_global_2017,grilo_roadkill_2020,Pinto2018,borda-de-agua_spatio-temporal_2011}.  

For this initial effort, we made a series of simplifying assumptions that facilitated obtaining exact analytical results. To simplify the link between the stochastic reaction-diffusion dynamics and the survival-problem approach and to make the stationary connection more explicit, we considered a zero-width limit of the road and subsumed collision rates and road width under a single traffic-intensity parameter, $\eta$. This simplification, however, yields accurate mortality estimates across a broad range of parameter values, with larger errors occurring in scenarios where intrinsic death rates are very high or roads are very far from the animal's home-range center. Importantly, in both of these scenarios, the impact of the road in an animal's lifetime is already very small, indicating that our vanishingly narrow road approximation is broadly applicable beyond the strict separation of scales we imposed in its derivation.

In an additional simplifying assumption, we considered scenarios where range-resident animals do not react to road presence. This assumption may hold for certain species, such as giant anteaters \citep{noonan_roads_2022} or jaguars \citep{cerqueira2021direct}, and for individuals that have not yet learned how to respond to roads. For example, young animals or populations newly exposed to roads. However, animals often alter their movement patterns near roads or vehicles in many other scenarios \citep{jacobson2016behavior,brieger2022behavioural,benard_integration_2024}. Introducing these behavioral responses complicates the calculation of collision times, making the general case mathematically intractable. However, in the limit where collisions can be treated as a stationary problem and mortality is determined primarily by the time animals spend in the road, such behavioral responses are already encoded in the observed individual range distribution. This enables extending our collision-time calculations to cases where animals exhibit more complex responses to roads leveraging already available home-range estimators \citep{calabrese2016}. More generally, if animals develop behavioral responses over time to reduce the likelihood of encounters, our baseline calculations provide an upper bound on mortality assuming maximally naïve individuals.

Similarly, our framework describes incoming traffic as a time point process, neglecting the time vehicles spend on the road, and treats roads as simple linear features. In the stationary regime, the effect of animal movement behavior on collision times is determined solely by the animal's range distribution evaluated at the road location. From an applied point of view, this result allows extending this model to more complex road networks, provided the integral of the animal range distribution over the network can still be computed. However, assuming that traffic is a point process neglects vehicle dynamics along the road. Instead, driver reactions, such as braking or perception-reaction delays, and external factors like speed limitations, are all aggregated into a phenomenological parameter, $q$, that represents the conditional probability of collision conditional on spatiotemporal animal-vehicle overlap. This approximation is essential to ensure mathematical tractability, but it eliminates many of the mechanistic processes responsible for collision outcomes. Relaxing this point-process assumption is challenging, and its validity should be tested through simulations that resolve vehicle trajectories \citep{Hels2001, benard_integration_2024}. Such trajectory-based approaches would allow these processes to be modeled more mechanistically, reducing the need to subsume them into $q$ and clarifying the conditions under which the point process approximation remains appropriate.

With the ongoing and rapid expansion of road networks worldwide \citep{meijer2018global,engert2024ghost}, reducing WVCs has become an increasingly important issue for both biodiversity conservation and transportation safety \citep{barbosa2020simulating,langbein2011traffic}, driving substantial investments in a range of mitigation strategies \citep{Glista2009, Denneboom2021}. However, at a more fundamental level, a theoretical framework untangling how the different processes underlying WVCs determine collision risk is lacking. Our framework serves as a key starting point for developing a predictive framework and a first step toward bridging the gap between road and movement ecology, providing deeper insights into the complex dynamics of WVC patterns. Taken as a baseline, our model suggests that within a plausible parameter range for medium and large range-resident organisms, WVCs cannot only significantly influence but, in some cases, dominate total mortality. Under these conditions, roadkill is not an additive source of mortality, but a primary demographic driver. Baselines such as this can help calibrate the expected effect of interventions, particularly in systems where carcass counts may fail to accurately estimate demographic impact if the demographic structure of a population is unknown. By identifying the key factors driving WVC patterns, our approach strengthens efforts in wildlife conservation and promotes sustainable transportation practices.

\section*{Acknowledgements}
This work was partially funded by the Center of Advanced Systems Understanding (CASUS), which is financed by Germany’s Federal Ministry of Research, Technology and Space (BMFTR) and by the Saxon Ministry for Science, Culture and Tourism (SMWK) with tax funds on the basis of the budget approved by the Saxon State Parliament. The authors also received partial support from the Simons Foundation through grant 284558FY19 (RMG); FAPESP through a BIOTA Jovem Pesquisador Grant 2019/05523-8 (RMG), ICTP-SAIFR 2021/14335-0 (RMG), and a Master's fellowship 2019/26736-0 (BGF); and Instituto Serrapilheira through grant Serra-1911-31200 (RMG, BGF). CHF was supported by NSF eMB 2527228.
        

\newpage
\newpage
\section*{Appendices}
\begin{appendices}

\section{Discretization and simulation setup.}\label{sec:micro}

We can measure collision times and subsequently validate analytical calculations by numerically simulating the stochastic reaction-diffusion process specified by our individual-based model. Trajectories of the diffusion process may be generated by integrating the Ornstein-Uhlenbeck equation \eqref{eq:OU} using a Milstein algorithm with discretized time steps $\Delta t$ \citep{toral_stochastic_2014}. The challenge then becomes how to count interactions, since in the point-like limit \eqref{eq:deltaapprox} the jump process lands on $d$ with null probability. Two routes are possible. One possibility is to explicitly simulate a finite-width domain with the corresponding Poisson process of possible interaction events over it. We instead pursue another, non-trivial solution, which is to assign a probability $P_\times$ of terminating the process at every time-step where the interacting line is crossed. We highlight this possibility for ecological applications, as effective probabilities per crossing might be an output of a fine-grained model of ballistic motion at the road scale \citep{Gibbs2002}, or a measurable quantity in the field \citep{Hels2001}.

Evaluating $P_\times$ as a function of $\eta$ is a delicate matter. This can be immediately seen from the fact that trajectories of the SDE are a.s. non-differentiable, meaning that the number of crossings diverges to infinity in the continuum time limit. The point-sink limit in \eqref{eq:deltaapprox}) and the continuum limit $\Delta t \rightarrow 0^+$ do not commute. Nonetheless, analysis from the related case of partially reflecting boundaries shows that the finite-width region may be parametrized as a boundary-layer of width proportional to $\sqrt{\Delta t}$. The full theory, developed in \citep{erban_reactive_2007}, shows that
\begin{equation}
    P_\times = \sqrt{\frac{\pi \Delta t}{2 \Sigma^2(d)}}\eta,
\end{equation}
where $\Sigma^2(d)$ is the infinitesimal covariance of the process at $d$. To obtain the statistics of the collision time $R$, we run $10^4$ independent realizations of the process in two spatial dimensions at a fixed model parameterization and obtained averages over the ensemble of realizations.

\section{Stationary mean first-passage time of the OU process}\label{app:smfpt}
A classical result in first passage theory is that the MGF of the first-passage time of an OU process conditioned on starting at $x_0$ follows \citep{siegert_first_1951}
\begin{equation}\langle e^{-sT}\rangle_{x_0} = 
    \begin{cases}
        \displaystyle e^{(x_0^2 - d^2)/4\sigma^2} \frac{D_{-s\tau}(-x_0/\sigma)}{D_{-s\tau}(-d/\sigma)}, & x_0 < a \\[1em]
        \displaystyle e^{(x_0^2 - d^2)/4\sigma^2} \frac{D_{-s\tau}(x_0/\sigma)}{D_{-s\tau}(d/\sigma)}, & x_0 \geq a,
    \end{cases}
\end{equation}
where $D_{-\nu}$ is the parabolic cylinder function with parameter $-\nu$. The stationary first passage time MGF then follows from the local-global correspondence \citep{pitman_lengths_1997, Figueiredo2025},
\begin{equation}
    \langle e^{-sT}\rangle = -\frac{\sigma^2 p_{\mathrm{st}}(d)}{s\tau}\frac{\mathrm{d}}{\mathrm{d}x_0}\langle e^{-sT}\rangle_{x_0} \bigg|^{d^+}_{d^{-}}.
\end{equation}
The identity above can be rewritten as
\begin{equation}
    \langle e^{-sT}\rangle = - \frac{e^{-d^2/2\sigma^2}}{\sqrt{2\pi}s\tau} \frac{\mathrm{d}}{\mathrm{d}x_0}\bigg |_{x_0 = d} e^{\frac{x_0^2 - d^2}{4\sigma^2}} \frac{D_{-s\tau}(x_0/\sigma)D_{-s\tau}(-d/\sigma) - D_{-s\tau}(-x_0/\sigma)D_{-s\tau}(d/\sigma)}{D_{-s\tau}(d/\sigma)D_{-s\tau}(-d/\sigma)},
\end{equation}
where the derivative on the right hand side is identical in value to the Wronskian of the parabolic cylinder equation, which evaluates to $-\sqrt{2\pi}/\sigma \Gamma(s\tau)$ \citep{borodin_handbook_2002}. Using $s\tau\Gamma(s\tau) = \Gamma(s\tau + 1)$ yields \eqref{eq:mgfstationaryOU}. 
 
To evaluate the first moment of this distribution, we consider that for $\text{Re}\;\nu > 0$, the parabolic cylinder function can be represented as the series \citep{borodin_handbook_2002},
\begin{align}
\begin{split}
    D_{-\nu}(x) &= e^{-x^2/4}2^{-\nu/2}\sqrt{\pi}\Bigg[\Gamma\left(\frac{\nu+1}{2}\right)^{-1}\left(1 + \sum_{k = 1}^\infty\frac{2^k(\nu/2)^{\overline{k-1}}}{(2k)!}x^{2k}\right)\\
    &-\Gamma\left(\frac{\nu}{2}\right)^{-1}\sqrt{2}x\left(1 + \sum_{k = 1}^\infty\frac{2^k((\nu+1)/2)^{\overline{k-1}}}{(2k+1)!}x^{2k}\right)\Bigg],
\end{split}
\end{align}
where $(a)^{\overline{k}}$ denotes the Pochhammer polynomial. In particular, $D_0(x) = e^{-x^2/4}$. Differentiating Eq.\,(\ref{eq:mgfstationaryOU}) with respect to $s$ at $s = 0$ one finds, from the definition of $\Gamma(x)$,
\begin{equation}
    -\frac{1}{\tau}\frac{\partial}{\partial s} \langle e^{-sT}\rangle\bigg |_{s=0^+} = \int_0^\infty e^{-z}\log(z)\text{d}z + e^{d^2/4\sigma^2}\frac{\partial}{\partial \nu}\left[D_{-\nu}\left(\frac{d}{\sigma}\right) + D_{-\nu}\left(-\frac{d}{\sigma}\right)\right]\Bigg |_{\nu = 0^+},
\end{equation}
The term in brackets selects only the even part of $D_{-\nu}$ in $d/\sigma$, and can be explicitly evaluated to be
\begin{align}
\begin{split}
    e^{d^2/4\sigma^2}\frac{\partial}{\partial \nu}\left[D_{-\nu}\left(\frac{d}{\sigma}\right) + D_{-\nu}\left(-\frac{d}{\sigma}\right)\right]\Bigg |_{\nu=0} &= -\log 2 - \frac{1}{\sqrt{\pi}}\int_0^\infty \frac{e^{-z}\log(z)}{\sqrt{z}}\text{d} z \\&+ 2 \sum_{k = 1}^\infty \frac{(2k-2)!!}{(2k)!}\left(\frac{d}{\sigma}\right)^{2k}.
\end{split}
\end{align}
In the last power series, the quotient of successive terms is a rational function of $k$, so it can be identified with a hypergeometric function in $(d/\sigma)^2$. Using the known identities,
\begin{align}
    \int_0^\infty e^{-z}\log(z)\text{d}z &= -\gamma, \\
    \int_0^\infty \frac{e^{-z}\log(z)}{\sqrt{z}}\text{d}z  &= -\sqrt{\pi}\left(\gamma + 2\log 2\right),
\end{align}
where $\gamma$ is the Euler-Mascheroni constant, we arrive (\ref{eq:stationaryMFPTOU}).

\section{Stationary moments of $K$} \label{sec:momentK}
To obtain the stationary moments of the collision time $R$, we need to obtain the moments of the first hitting time $T$, and the killing time assuming an initial condition at the road $K$. We obtained the stationary first moment of $T$ in Appendix\,\ref{app:smfpt}. The MGF of $K$ can be expressed in terms of the stationary MGF of $T$, so the explicit evaluation of $\langle K^n \rangle$ reduces to a power series composition \citep{Figueiredo2025}. Using Bell polynomials $\mathcal B_{n,k}$ \citep{osullivan_moivre_2022}, one can write
\begin{equation}
\begin{split}
    \langle K^n\rangle &= \sum_{k=0}^n \frac{k!}{\omega_d^k}\mathcal B_{n,k}\left(1, 2\langle T \rangle,\cdots,(n-k+1)\langle T^{n - k}\rangle \right).
\end{split}
\end{equation}

We can see, therefore, that the $n$-th moment of $K$ depends only on the $(n-1)$ first stationary moments of $T$. In particular, we find a universal relationship between the mean collision time and the stationary probability distribution coinciding with the aforementioned stationary sampling limit, as the first few terms of the expansion above can be evaluated to
\begin{align}
    \langle K \rangle &= \omega_d^{-1}, \label{eq:meanK}\\
    \langle K^2 \rangle &= 2\omega_d^{-2} + 2\langle T \rangle \omega_d^{-1}, \\
    \langle K^3 \rangle &= 6\omega_d^{-3} + 12\langle T \rangle \omega_d^{-2} + 3\langle T^2\rangle \omega_d^{-1}.
\end{align}
Note the highest degree monomial in $\omega_d^{-1}$ coincides with the corresponding moment of the exponential distribution.

\section{Effect of road width}\label{app:roadwidth}

Here we extend the model to roads with finite width to validate the narrow-road approximation used in the main text and to account for cases where road width is comparable to the home range (e.g., for smaller organisms). This scenario is represented by the killing rate on the LHS of \eqref{eq:deltaapprox}. Keeping the assumption that the animal basal lifetime is an exponentially distributed random variable, $\Delta\sim \mathrm{Exp}(\delta)$, the probability of premature death is
\begin{equation}\label{eq:finitewidth}
    \left[\hat L^\dagger_{x_0} - \delta - \omega(x_0)\right]\langle e^{-\delta R}\rangle_{x_0} = -\omega(x_0),
\end{equation}
such that $\lim_{x_0 \rightarrow\pm \infty} \langle e^{-\delta R}\rangle_{x_0} = 0$. Since $\omega(x_0)$ is piecewise constant, the functional form of the fundamental solutions of the homogeneous equation are the same as in the 
zero-width limit, in terms of parabolic cylinder functions \citep{borodin_handbook_2002}. In the interior of the road, $x\in[d-\Delta d/2, d + \Delta d/2]$, a particular solution of the inhomogeneous system is the moment generating function for an exponential random time with rate $\nu$, which does not depend on $x_0$. Thus, the finite-width solution can be constructed, up to four constants, as
\begin{equation}\label{eq:exact}
    \langle e^{-\delta R} \rangle_{x_0} = \begin{cases}
        C_1 e^{\frac{x_0^2}{4\sigma^2}} D_{-\delta\tau}\left(-\frac{x_0}{\sigma}\right),&x_0 < d-\Delta d/2, \\
        \frac{\nu}{\delta+\nu} + C_2 e^{\frac{x_0^2}{4\sigma^2}} D_{-(\delta+\nu)\tau}\left(-\frac{x_0}{\sigma}\right) + C_3 e^{\frac{x_0^2}{4\sigma^2}} D_{-(\delta+\nu)\tau}\left(\frac{x_0}{\sigma}\right), & |x _0 - d| \leq \Delta d/2, \\
        C_4 e^{\frac{x_0^2}{4\sigma^2}} D_{-\delta\tau}\left(\frac{x_0}{\sigma}\right), & x_0 > d+\Delta d/2.
    \end{cases}
\end{equation}
Imposing continuous differentiability of $\langle e^{-\delta R} \rangle_{x_0}$ at $x_0 = d\pm \Delta d/2$ creates a $4\times 4$ linear system in $(C_1,C_2,C_3,C_4)$ which uniquely fixes the solution (detailed in Appendix \ref{app:finitewidth}). The average of this MGF over the stationary distribution can be calculated by exploiting once again the local-global correspondence \citep{Figueiredo2025}. Integrating both sides of \eqref{eq:finitewidth} multiplied by the stationary probability $p_\mathrm{st}(x_0)$ gives
\begin{equation}
    \langle e^{-\delta R} \rangle = \left[\frac{\nu}{2(\delta + \nu)}\mathrm{erf}\left(\frac{x_0}{\sqrt{2} \sigma}\right)- \frac{1}{{\sqrt{2\pi}}}\frac{\sigma \nu}{\tau \delta(\delta + \nu)}e^{-\frac{x_0^2}{2\sigma^2}} \frac{\mathrm{d}}{\mathrm{d}x_0}\langle e^{-\delta R}\rangle_{x_0}\right]_{x_0 = d - \Delta d/2}^{x_0 = d + \Delta d/2}.
\end{equation}
A useful limit of this result arises in the case where the scales of motion are much smaller than the width of the road, so that the road may be approximated by the interval $[d-\Delta d/2, \infty)$, and the expressions above simplify to
\begin{align}
\begin{split}
    \langle e^{-\delta R} \rangle &\rightarrow \frac{\nu}{2(\delta + \nu)} \mathrm{erfc}\left(\frac{d - \Delta d/2}{\sqrt{2} \sigma}\right) \\ &\quad+ \frac{1}{{\sqrt{2\pi}}}\frac{\sigma \nu^2}{\tau \delta(\delta + \nu)^2}e^{-\frac{x_0^2}{2\sigma^2}} \\ &\quad\times\Bigg[
\frac{
D_{-\tau\delta}\!\left(-\frac{d-\Delta d/2}{\sigma}\right)
}{
\frac{d-\Delta d/2}{\sigma}\,
D_{-\tau\delta}\!\left(-\frac{d-\Delta d/2}{\sigma}\right)
+
D_{1-\tau\delta}\!\left(-\frac{d-\Delta d/2}{\sigma}\right)
}
\\ &\quad-
\frac{
D_{-\tau(\delta+\nu)}\!\left(\frac{d-\Delta d/2}{\sigma}\right)
}{
\frac{d-\Delta d/2}{\sigma}\,
D_{-\tau(\delta+\nu)}\!\left(\frac{d-\Delta d/2}{\sigma}\right)
-
D_{1-\tau(\delta+\nu)}\!\left(\frac{d-\Delta d/2}{\sigma}\right)
}
\Bigg]^{-1}.
\end{split}
\end{align}

\noindent This new approximation, the exact solution in \eqref{eq:exact} and the zero-width limit considered in the main text define three treatments of the problem. Each of them provides different mortality estimates: $\langle e^{-\delta R} \rangle_{\mathrm{Exact}}$ for the full solution, $ \langle e^{-\delta R} \rangle_{\mathrm{Half-infinite}}$ for the half-infinite road limit, and the main-text treatment $\langle e^{-\delta R}\rangle_\mathrm{Zero-width}$ where the road width is approximated by a  $\delta$ function. By construction, the half-infinite approximation systematically overestimates mortality, because it assumes a fixed, larger domain where collisions are possible. The zero-width approximation can either over or underestimate mortality, depending on whether the local time density of the trajectory at the center of the road domain is higher or lower than its average over the whole road with finite width. 

\begin{figure}[H]
    \centering
    \includegraphics[width=\linewidth]{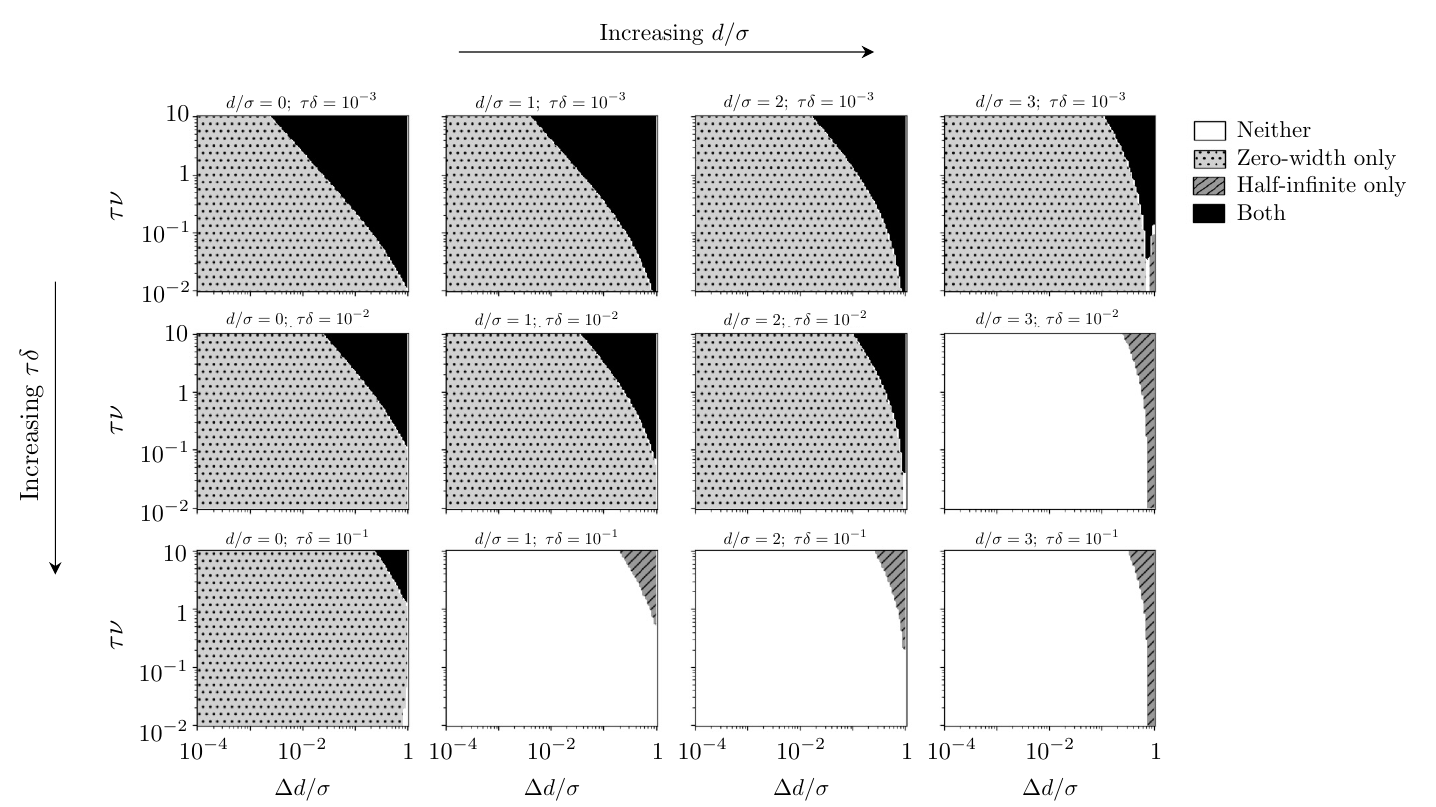}
    \caption{Parameter regions where the zero-width and half-line approximations reproduce the exact stationary MGF value within 10\% relative error. Dotted, hashed, and black regions indicate where the zero-width approximation, half-infinite road approximation, or both are accurate, respectively; white regions indicate where neither approximation is accurate.}
    \label{fig:SIFig1}
\end{figure}

To compare the three approximations over a variety of regimes, we define a relative error in the estimate of excess mortality
\begin{equation}
    \epsilon_i = \frac{\rvert \langle e^{-sR\rangle} \rangle_\mathrm{i} - \langle e^{-sR}\rangle_\mathrm{Exact}\rvert}{\langle e^{-sR}\rangle_\mathrm{Exact}},
\end{equation}
where $i =$Zero-width, Half-infinite. Assuming an acceptable relative error of $10\%$, we analyze the adequacy of each approximation within an broad and biologically plausible parameter sweep (Fig.\,\ref{fig:SIFig1}).
We find that the zero-width approximation deviates significantly only in extreme scenarios where intrinsic death rates are very large, or the road lies in the far tail of the animal's stationary space utilization function, with road width having a minor effect. Importantly, in both of these scenarios, the relative impact of the road on lifespan, as obtained from the exact solution, is already small. Finally, we observe that the half-infinite approximation is often adequate when road width becomes comparable to home-range dimensions.

\section{Non-zero width solution}\label{app:finitewidth}
Imposing continuity of both $\langle e^{-\delta R}\rangle_{d \pm \Delta d/2}$ and $\frac{\mathrm{d}}{\mathrm{d}x_0}\langle e^{-\delta R}\rangle_{d \pm \Delta d/2}\Big |_{x_0 = d \pm \Delta d/2}$, and exploiting identities of the parabolic cylinder function \citep{borodin_handbook_2002}, we find, after simplification, that the constants $(C_1,C_2,C_3,C_4)$ are determined by as the unique solution to the linear system
\begin{equation}
\begin{pmatrix}
M_{11} & M_{12} & M_{13} & M_{14} \\
M_{21} & M_{22} & M_{23} & M_{24} \\
M_{31} & M_{32} & M_{33} & M_{34} \\
M_{41} & M_{42} & M_{43} & M_{44}
\end{pmatrix}
\begin{pmatrix}
C_1\\ C_2\\ C_3\\ C_4
\end{pmatrix}
=
\frac{\nu}{\delta+\nu}
\begin{pmatrix}
\exp\left(-\frac{(d - \Delta d/2)^{2}}{4}\right)\\[2pt]
0\\[2pt]
-\exp\left(-\frac{(d + \Delta d/2)^{2}}{4}\right)\\[2pt]
0
\end{pmatrix},
\end{equation}
where
\begin{align}
M_{11} &= D_{-\tau \delta}\!\left(-\frac{d-\Delta d/2}{\sigma}\right), \\
M_{12} &= -D_{-\tau(\delta+\nu)}\!\left(-\frac{d-\Delta d/2}{\sigma}\right), \\
M_{13} &= -D_{-\tau(\delta+\nu)}\!\left(\frac{d-\Delta d/2}{\sigma}\right), \\
M_{14} &= 0, \\[6pt]
M_{21} &= \frac{d-\Delta d/2}{\sigma}\,
         D_{-\tau \delta}\!\left(-\frac{d-\Delta d/2}{\sigma}\right)
         +D_{1-\tau \delta}\!\left(-\frac{d-\Delta d/2}{\sigma}\right), \\
M_{22} &= -\Bigg(
         \frac{d-\Delta d/2}{\sigma}\,
         D_{-\tau(\delta+\nu)}\!\left(-\frac{d-\Delta d/2}{\sigma}\right)
         +D_{1-\tau(\delta+\nu)}\!\left(-\frac{d-\Delta d/2}{\sigma}\right)
         \Bigg), \\
M_{23} &= -\Bigg(
         \frac{d-\Delta d/2}{\sigma}\,
         D_{-\tau(\delta+\nu)}\!\left(\frac{d-\Delta d/2}{\sigma}\right)
         -D_{1-\tau(\delta+\nu)}\!\left(\frac{d-\Delta d/2}{\sigma}\right)
         \Bigg), \\
M_{24} &= 0, \\[6pt]
M_{31} &= 0, \\
M_{32} &= D_{-\tau(\delta+\nu)}\!\left(-\frac{d+\Delta d/2}{\sigma}\right), \\
M_{33} &= D_{-\tau(\delta+\nu)}\!\left(\frac{d+\Delta d/2}{\sigma}\right), \\
M_{34} &= -D_{-\tau \delta}\!\left(\frac{d+\Delta d/2}{\sigma}\right), \\[6pt]
M_{41} &= 0, \\
M_{42} &= \frac{d+\Delta d/2}{\sigma}\,
         D_{-\tau(\delta+\nu)}\!\left(-\frac{d+\Delta d/2}{\sigma}\right)
         +D_{1-\tau(\delta+\nu)}\!\left(-\frac{d+\Delta d/2}{\sigma}\right), \\
M_{43} &= \frac{d+\Delta d/2}{\sigma}\,
         D_{-\tau(\delta+\nu)}\!\left(\frac{d+\Delta d/2}{\sigma}\right)
         -D_{1-\tau(\delta+\nu)}\!\left(\frac{d+\Delta d/2}{\sigma}\right), \\
M_{44} &= -\Bigg(
         \frac{d+\Delta d/2}{\sigma}\,
         D_{-\tau \delta}\!\left(\frac{d+\Delta d/2}{\sigma}\right)
         -D_{1-\tau \delta}\!\left(\frac{d+\Delta d/2}{\sigma}\right)
         \Bigg).
\end{align}
\end{appendices}

\begin{thebibliography}{84}
\providecommand{\natexlab}[1]{#1}
\providecommand{\url}[1]{\texttt{#1}}
\expandafter\ifx\csname urlstyle\endcsname\relax
  \providecommand{\doi}[1]{doi: #1}\else
  \providecommand{\doi}{doi: \begingroup \urlstyle{rm}\Url}\fi

\bibitem[Abra et~al.(2019)Abra, Granziera, Huijser, Ferraz, Haddad, and
  Paolino]{abra2019pay}
Fernanda~Delborgo Abra, Beatriz~Machado Granziera, Marcel~Pieter Huijser, Katia
  Maria Paschoaletto Micchi de~Barros Ferraz, Camilla~Mansur Haddad, and
  Roberta~Montanheiro Paolino.
\newblock Pay or prevent? human safety, costs to society and legal perspectives
  on animal-vehicle collisions in s{\~a}o paulo state, brazil.
\newblock \emph{Plos One}, 14\penalty0 (4):\penalty0 e0215152, 2019.
\newblock \doi{10.1371/journal.pone.0215152}.

\bibitem[Al~Shimemeri and Arabi(2013)]{al2013review}
Abdullah Al~Shimemeri and Yaseen Arabi.
\newblock A review of large animal vehicle accidents with special focus on
  arabian camels.
\newblock \emph{Journal of Emergency Medicine, Trauma \& Acute Care},
  2012\penalty0 (1):\penalty0 21, 2013.
\newblock \doi{10.5339/jemtac.2012.21}.

\bibitem[Alili et~al.(2005)Alili, Patie, and
  Pedersen]{alili_representations_2005}
L.~Alili, P.~Patie, and J.~L. Pedersen.
\newblock Representations of the {First} {Hitting} {Time} {Density} of an
  {Ornstein}-{Uhlenbeck} {Process}.
\newblock \emph{Stochastic Models}, 21\penalty0 (4):\penalty0 967--980, October
  2005.
\newblock ISSN 1532-6349.
\newblock \doi{10.1080/15326340500294702}.
\newblock URL \url{https://doi.org/10.1080/15326340500294702}.
\newblock Publisher: Taylor \& Francis \_eprint:
  https://doi.org/10.1080/15326340500294702.

\bibitem[Anderson and May(1991)]{Anderson1991}
Roy~M. Anderson and Robert~M. May.
\newblock \emph{Infectious Diseases of Humans: Dynamics and Control}.
\newblock Oxford University Press, 1991.
\newblock \doi{10.1093/oso/9780198545996.001.0001}.

\bibitem[Ascensão et~al.(2019)Ascensão, D'Amico, and
  Barrientos]{ascensao_validation_2019}
Fernando Ascensão, Marcello D'Amico, and Rafael Barrientos.
\newblock Validation data is needed to support modelling in {Road} {Ecology}.
\newblock \emph{Biological Conservation}, 230:\penalty0 199--200, February
  2019.
\newblock ISSN 00063207.
\newblock \doi{10.1016/j.biocon.2018.12.023}.
\newblock URL
  \url{https://linkinghub.elsevier.com/retrieve/pii/S0006320718316343}.

\bibitem[Aujla et~al.(2022)Aujla, Montoya, Montoya, Rea, and
  Hesse]{aujla2022health}
Braedon Aujla, David Montoya, Chris Montoya, Roy~V Rea, and Gayle Hesse.
\newblock Health care access and injury patterns in patients following
  moose-and deer-vehicle collisions in north-central british columbia.
\newblock \emph{British Columbia Medical Journal}, 64\penalty0 (7):\penalty0
  292--296, 2022.

\bibitem[Banerjee et~al.(2020)Banerjee, Duflo, and Qian]{banerjee2020}
Abhijit Banerjee, Esther Duflo, and Nancy Qian.
\newblock On the road: Access to transportation infrastructure and economic
  growth in china.
\newblock \emph{Journal of Development Economics}, 145:\penalty0 102442, 2020.
\newblock \doi{10.1016/j.jdeveco.2020.102442}.

\bibitem[Barbosa et~al.(2020)Barbosa, Schumaker, Brandon, Bager, and
  Grilo]{barbosa2020simulating}
Priscilla Barbosa, Nathan~H Schumaker, Kristin~R Brandon, Alex Bager, and Clara
  Grilo.
\newblock Simulating the consequences of roads for wildlife population
  dynamics.
\newblock \emph{Landscape and urban planning}, 193:\penalty0 103672, 2020.
\newblock \doi{10.1016/j.landurbplan.2019.103672}.

\bibitem[Blackwell et~al.(2016)Blackwell, DeVault, Fern{\'a}ndez-Juricic, Gese,
  Gilbert-Norton, and Breck]{Blackwell2016}
Bradley~F Blackwell, Travis~L DeVault, Esteban Fern{\'a}ndez-Juricic, Eric~M
  Gese, Lynne Gilbert-Norton, and Stewart~W Breck.
\newblock No single solution: application of behavioural principles in
  mitigating human--wildlife conflict.
\newblock \emph{Animal Behaviour}, 120:\penalty0 245--254, 2016.
\newblock \doi{10.1016/j.anbehav.2016.07.013}.

\bibitem[Borda-de Água et~al.(2011)Borda-de Água, Navarro, Gavinhos, and
  Pereira]{borda-de-agua_spatio-temporal_2011}
Luís Borda-de Água, Laetitia Navarro, Catarina Gavinhos, and Henrique~M.
  Pereira.
\newblock Spatio-temporal impacts of roads on the persistence of populations:
  analytic and numerical approaches.
\newblock \emph{Landscape Ecology}, 26\penalty0 (2):\penalty0 253--265,
  February 2011.
\newblock ISSN 0921-2973, 1572-9761.
\newblock \doi{10.1007/s10980-010-9546-2}.
\newblock URL \url{http://link.springer.com/10.1007/s10980-010-9546-2}.

\bibitem[Borodin and Salminen(2002)]{borodin_handbook_2002}
Andrei~N. Borodin and Paavo Salminen.
\newblock \emph{Handbook of {Brownian} {Motion} - {Facts} and {Formulae}}.
\newblock Probability and {Its} {Applications}. Birkhäuser, Basel, 2002.
\newblock ISBN 978-3-7643-6705-3 978-3-0348-8163-0.
\newblock \doi{10.1007/978-3-0348-8163-0}.
\newblock URL \url{http://link.springer.com/10.1007/978-3-0348-8163-0}.

\bibitem[Brieger et~al.(2022)Brieger, K{\"a}mmerle, Hagen, and
  Suchant]{brieger2022behavioural}
Falko Brieger, Jim-Lino K{\"a}mmerle, Robert Hagen, and Rudi Suchant.
\newblock Behavioural reactions to oncoming vehicles as a crucial aspect of
  wildlife-vehicle collision risk in three common wildlife species.
\newblock \emph{Accident Analysis \& Prevention}, 168:\penalty0 106564, 2022.
\newblock \doi{10.1016/j.aap.2021.106564}.

\bibitem[Bénard et~al.(2024)Bénard, Lengagne, and
  Bonenfant]{benard_integration_2024}
Annaëlle Bénard, Thierry Lengagne, and Christophe Bonenfant.
\newblock Integration of animal movement into wildlife-vehicle collision
  models.
\newblock \emph{Ecological Modelling}, 492:\penalty0 110690, June 2024.
\newblock ISSN 03043800.
\newblock \doi{10.1016/j.ecolmodel.2024.110690}.
\newblock URL
  \url{https://linkinghub.elsevier.com/retrieve/pii/S0304380024000784}.

\bibitem[Calabrese et~al.(2016)Calabrese, Fleming, and Gurarie]{calabrese2016}
Justin~M Calabrese, Chris~H Fleming, and Eliezer Gurarie.
\newblock ctmm: an r package for analyzing animal relocation data as a
  continuous-time stochastic process.
\newblock \emph{Methods in Ecology and Evolution}, 7\penalty0 (9):\penalty0
  1124--1132, 2016.
\newblock \doi{10.1111/2041-210X.12559}.

\bibitem[Caughley(1966)]{caughley1966}
Graeme Caughley.
\newblock Mortality patterns in mammals.
\newblock \emph{Ecology}, 47\penalty0 (6):\penalty0 906--918, 1966.
\newblock \doi{10.2307/1935638}.

\bibitem[Ceia‐Hasse et~al.(2017)Ceia‐Hasse, Borda‐de‐Água, Grilo, and
  Pereira]{ceiahasse_global_2017}
Ana Ceia‐Hasse, Luís Borda‐de‐Água, Clara Grilo, and Henrique~M.
  Pereira.
\newblock Global exposure of carnivores to roads.
\newblock \emph{Global Ecology and Biogeography}, 26\penalty0 (5):\penalty0
  592--600, May 2017.
\newblock ISSN 1466-822X, 1466-8238.
\newblock \doi{10.1111/geb.12564}.
\newblock URL \url{https://onlinelibrary.wiley.com/doi/10.1111/geb.12564}.

\bibitem[Cerqueira et~al.(2021)Cerqueira, de~Rivera, Jaeger, and
  Grilo]{cerqueira2021direct}
Rafaela~Cobucci Cerqueira, Oscar~Rodr{\'\i}guez de~Rivera, Jochen~AG Jaeger,
  and Clara Grilo.
\newblock Direct and indirect effects of roads on space use by jaguars in
  brazil.
\newblock \emph{Scientific reports}, 11\penalty0 (1):\penalty0 22617, 2021.

\bibitem[Conover(2019)]{conover2019numbers}
Michael~R Conover.
\newblock Numbers of human fatalities, injuries, and illnesses in the united
  states due to wildlife.
\newblock \emph{Human-Wildlife Interactions}, 13\penalty0 (2):\penalty0
  264--276, 2019.
\newblock \doi{10.26077/r59n-bv76}.

\bibitem[da~Rosa and Bager(2013)]{da2013review}
Clarissa~Alves da~Rosa and Alex Bager.
\newblock Review of the factors underlying the mechanisms and effects of roads
  on vertebrates.
\newblock \emph{Oecologia Australis}, 17\penalty0 (1):\penalty0 6--19, 2013.
\newblock \doi{10.4257/oeco.2013.1701.02}.

\bibitem[Das et~al.(2023)Das, Kenkre, Nathan, and
  Giuggioli]{das_misconceptions_2023}
Debraj Das, V.~M. Kenkre, Ran Nathan, and Luca Giuggioli.
\newblock Misconceptions about quantifying animal encounter and interaction
  processes.
\newblock \emph{Frontiers in Ecology and Evolution}, 11, 2023.
\newblock ISSN 2296-701X.
\newblock \doi{10.3389/fevo.2023.1230890}.
\newblock URL
  \url{https://www.frontiersin.org/articles/10.3389/fevo.2023.1230890}.

\bibitem[Denneboom et~al.(2021)Denneboom, Bar-Massada, and
  Shwartz]{Denneboom2021}
Dror Denneboom, Avi Bar-Massada, and Assaf Shwartz.
\newblock Factors affecting usage of crossing structures by wildlife – {A}
  systematic review and meta-analysis.
\newblock \emph{Science of The Total Environment}, 777:\penalty0 146061, July
  2021.
\newblock ISSN 00489697.
\newblock \doi{10.1016/j.scitotenv.2021.146061}.
\newblock URL
  \url{https://linkinghub.elsevier.com/retrieve/pii/S0048969721011281}.

\bibitem[Engert et~al.(2024)Engert, Campbell, Cinner, Ishida, Sloan, Supriatna,
  Alamgir, Cislowski, and Laurance]{engert2024ghost}
Jayden~E Engert, Mason~J Campbell, Joshua~E Cinner, Yoko Ishida, Sean Sloan,
  Jatna Supriatna, Mohammed Alamgir, Jaime Cislowski, and William~F Laurance.
\newblock Ghost roads and the destruction of asia-pacific tropical forests.
\newblock \emph{Nature}, 629\penalty0 (8011):\penalty0 370--375, 2024.
\newblock \doi{10.1038/s41586-024-07303-5}.

\bibitem[Erban and Chapman(2007)]{erban_reactive_2007}
Radek Erban and S.~Jonathan Chapman.
\newblock Reactive boundary conditions for stochastic simulations of
  reaction–diffusion processes.
\newblock \emph{Physical Biology}, 4\penalty0 (1):\penalty0 16--28, February
  2007.
\newblock ISSN 1478-3975.
\newblock \doi{10.1088/1478-3975/4/1/003}.
\newblock URL \url{https://doi.org/10.1088/1478-3975/4/1/003}.
\newblock Publisher: IOP Publishing.

\bibitem[Erban and Chapman(2020)]{erban2020}
Radek Erban and S~Jonathan Chapman.
\newblock \emph{Stochastic modelling of reaction--diffusion processes},
  volume~60.
\newblock Cambridge University Press, 2020.
\newblock \doi{10.1017/9781108628389}.

\bibitem[Fagan et~al.(2024)Fagan, Krishnan, Liao, Fleming, Liao, Lamb,
  Patterson, Wheeldon, Martinez-Garcia, Menezes, et~al.]{fagan2024}
William~F Fagan, Ananke Krishnan, Qianru Liao, Christen~H Fleming, Daisy Liao,
  Clayton Lamb, Brent Patterson, Tyler Wheeldon, Ricardo Martinez-Garcia,
  Jorge~FS Menezes, et~al.
\newblock Intraspecific encounters can lead to reduced range overlap.
\newblock \emph{Movement ecology}, 12\penalty0 (1):\penalty0 58, 2024.
\newblock \doi{10.1186/s40462-024-00501-w}.

\bibitem[Fleming et~al.(2014)Fleming, Calabrese, Mueller, Olson, Leimgruber,
  and Fagan]{fleming_fine-scale_2014}
Chris~H. Fleming, Justin~M. Calabrese, Thomas Mueller, Kirk~A. Olson, Peter
  Leimgruber, and William~F. Fagan.
\newblock From {Fine}-{Scale} {Foraging} to {Home} {Ranges}: {A} {Semivariance}
  {Approach} to {Identifying} {Movement} {Modes} across {Spatiotemporal}
  {Scales}.
\newblock \emph{The American Naturalist}, 183\penalty0 (5):\penalty0
  E154--E167, May 2014.
\newblock ISSN 0003-0147.
\newblock \doi{10.1086/675504}.
\newblock URL \url{https://www.journals.uchicago.edu/doi/full/10.1086/675504}.
\newblock Publisher: The University of Chicago Press.

\bibitem[Garcia~de Figueiredo et~al.(2025)Garcia~de Figueiredo, Calabrese,
  Fagan, and Martinez-Garcia]{Figueiredo2025}
Benjamin Garcia~de Figueiredo, Justin~M. Calabrese, William~F. Fagan, and
  Ricardo Martinez-Garcia.
\newblock Ergodicity shapes inference in biological reactions driven by a
  latent trajectory.
\newblock \emph{arXiv}, April 2025.
\newblock \doi{10.48550/arXiv.2409.11433}.
\newblock URL \url{http://arxiv.org/abs/2409.11433}.
\newblock arXiv:2409.11433 [cond-mat].

\bibitem[Gardiner(2009)]{gardiner_stochastic_2009}
Crispin Gardiner.
\newblock \emph{Stochastic {Methods}: {A} {Handbook} for the {Natural} and
  {Social} {Sciences}: 13}.
\newblock Springer, Berlin, 4th edition, January 2009.
\newblock ISBN 978-3-540-70712-7.

\bibitem[Gibbs and Shriver(2002)]{Gibbs2002}
James~P. Gibbs and W.~Gregory Shriver.
\newblock Estimating the {Effects} of {Road} {Mortality} on {Turtle}
  {Populations}.
\newblock \emph{Conservation Biology}, 16\penalty0 (6):\penalty0 1647--1652,
  2002.
\newblock ISSN 1523-1739.
\newblock \doi{10.1046/j.1523-1739.2002.01215.x}.
\newblock URL
  \url{https://onlinelibrary.wiley.com/doi/abs/10.1046/j.1523-1739.2002.01215.x}.

\bibitem[Glista et~al.(2009)Glista, DeVault, and DeWoody]{Glista2009}
David~J. Glista, Travis~L. DeVault, and J.~Andrew DeWoody.
\newblock A review of mitigation measures for reducing wildlife mortality on
  roadways.
\newblock \emph{Landscape and Urban Planning}, 91\penalty0 (1):\penalty0 1--7,
  May 2009.
\newblock ISSN 01692046.
\newblock \doi{10.1016/j.landurbplan.2008.11.001}.
\newblock URL
  \url{https://linkinghub.elsevier.com/retrieve/pii/S0169204608001886}.

\bibitem[Grilo et~al.(2010)Grilo, Bissonette, and Cramer]{Grilo2010}
C~Grilo, John~A Bissonette, and PA~Cramer.
\newblock \emph{Mitigation Measures to Reduce Impacts on Biodiversity}, pages
  73--114.
\newblock Nova Science Publishers, Inc., Hauppauge, NY, 2010.

\bibitem[Grilo et~al.(2020)Grilo, Koroleva, Andrášik, Bíl, and
  González‐Suárez]{grilo_roadkill_2020}
Clara Grilo, Elena Koroleva, Richard Andrášik, Michal Bíl, and Manuela
  González‐Suárez.
\newblock Roadkill risk and population vulnerability in {European} birds and
  mammals.
\newblock \emph{Frontiers in Ecology and the Environment}, 18\penalty0
  (6):\penalty0 323--328, August 2020.
\newblock ISSN 1540-9295, 1540-9309.
\newblock \doi{10.1002/fee.2216}.
\newblock URL
  \url{https://esajournals.onlinelibrary.wiley.com/doi/10.1002/fee.2216}.

\bibitem[Grilo et~al.(2021)Grilo, Borda‐de‐Água, Beja, Goolsby, Soanes,
  Le~Roux, Koroleva, Ferreira, Gagné, Wang, and
  González‐Suárez]{grilo_conservation_2021}
Clara Grilo, Luis Borda‐de‐Água, Pedro Beja, Eric Goolsby, Kylie Soanes,
  Aliza Le~Roux, Elena Koroleva, Flávio~Z. Ferreira, Sara~A. Gagné, Yun Wang,
  and Manuela González‐Suárez.
\newblock Conservation threats from roadkill in the global road network.
\newblock \emph{Global Ecology and Biogeography}, 30\penalty0 (11):\penalty0
  2200--2210, November 2021.
\newblock ISSN 1466-822X, 1466-8238.
\newblock \doi{10.1111/geb.13375}.
\newblock URL \url{https://onlinelibrary.wiley.com/doi/10.1111/geb.13375}.

\bibitem[Gunson and Teixeira(2015)]{gunson2015road}
Kari Gunson and Fernanda~Zimmermann Teixeira.
\newblock Road--wildlife mitigation planning can be improved by identifying the
  patterns and processes associated with wildlife-vehicle collisions.
\newblock \emph{Handbook of road ecology}, pages 101--109, 2015.
\newblock \doi{10.1002/9781118568170.ch13}.

\bibitem[Gunson et~al.(2011)Gunson, Mountrakis, and
  Quackenbush]{gunson_spatial_2011}
Kari~E. Gunson, Giorgos Mountrakis, and Lindi~J. Quackenbush.
\newblock Spatial wildlife-vehicle collision models: {A} review of current work
  and its application to transportation mitigation projects.
\newblock \emph{Journal of Environmental Management}, 92\penalty0 (4):\penalty0
  1074--1082, April 2011.
\newblock ISSN 03014797.
\newblock \doi{10.1016/j.jenvman.2010.11.027}.
\newblock URL
  \url{https://linkinghub.elsevier.com/retrieve/pii/S0301479710004305}.

\bibitem[Gurarie and Ovaskainen(2013)]{gurarie2013}
Eliezer Gurarie and Otso Ovaskainen.
\newblock Towards a general formalization of encounter rates in ecology.
\newblock \emph{Theoretical ecology}, 6:\penalty0 189--202, 2013.
\newblock \doi{10.1007/s12080-012-0170-4}.

\bibitem[Hartich and Godec(2019)]{hartich_interlacing_2019}
David Hartich and Aljaž Godec.
\newblock Interlacing relaxation and first-passage phenomena in reversible
  discrete and continuous space {Markovian} dynamics.
\newblock \emph{Journal of Statistical Mechanics: Theory and Experiment},
  2019\penalty0 (2):\penalty0 024002, February 2019.
\newblock ISSN 1742-5468.
\newblock \doi{10.1088/1742-5468/ab00df}.
\newblock URL \url{https://dx.doi.org/10.1088/1742-5468/ab00df}.
\newblock Publisher: IOP Publishing and SISSA.

\bibitem[Hels and Buchwald(2001)]{Hels2001}
Tove Hels and Erik Buchwald.
\newblock The effect of road kills on amphibian populations.
\newblock \emph{Biological Conservation}, 2001.
\newblock \doi{10.1016/S0006-3207(00)00215-9}.

\bibitem[Hothorn et~al.(2012)Hothorn, Brandl, and Müller]{Hothorn2012}
Torsten Hothorn, Roland Brandl, and Jörg Müller.
\newblock Large-{Scale} {Model}-{Based} {Assessment} of {Deer}-{Vehicle}
  {Collision} {Risk}.
\newblock \emph{PLoS ONE}, 7\penalty0 (2):\penalty0 e29510, February 2012.
\newblock ISSN 1932-6203.
\newblock \doi{10.1371/journal.pone.0029510}.
\newblock URL \url{https://dx.plos.org/10.1371/journal.pone.0029510}.

\bibitem[Hubbell(2001)]{Hubbell2001}
Stephen~P. Hubbell.
\newblock \emph{The Unified Neutral Theory of Biodiversity and Biogeography}.
\newblock Princeton University Press, 2001.
\newblock \doi{10.1890/04-0808}.

\bibitem[Huijser et~al.(2008)Huijser, McGowen, Fuller, Hardy, and
  Kocjolek]{huijser2008wildlife}
M~Huijser, PT~McGowen, J~Fuller, A~Hardy, and A~Kocjolek.
\newblock Wildlife-vehicle collision reduction study: report to congress (no.
  fhwa-hrt-08-034).
\newblock \emph{Washington, DC: US Department of Transportation}, 2008.

\bibitem[Itô(1969)]{ito_poisson_1969}
Kiyosi Itô.
\newblock \emph{Poisson {Point} {Processes} and {Their} {Application} to
  {Markov} {Processes}}.
\newblock Springer, 1969.
\newblock ISBN 978-981-10-0272-4.

\bibitem[Itô(1971)]{ito_poisson_1971}
Kiyosi Itô.
\newblock Poisson point processes attached to {Markov} processes.
\newblock \emph{Proceedings of the Sixth Berkeley Symposium on Mathematical
  Statistics and Probability (June 21-July 18, 1970, April 9-12, June 16-21 and
  July 19-22, 1971)}, 1971.
\newblock \doi{10.1525/9780520375918-015}.
\newblock Num Pages: 225.

\bibitem[Jacobson et~al.(2016)Jacobson, Bliss-Ketchum, de~Rivera, and
  Smith]{jacobson2016behavior}
Sandra~L Jacobson, Leslie~L Bliss-Ketchum, Catherine~E de~Rivera, and Winston~P
  Smith.
\newblock A behavior-based framework for assessing barrier effects to wildlife
  from vehicle traffic volume.
\newblock \emph{Ecosphere}, 7\penalty0 (4):\penalty0 e01345, 2016.
\newblock \doi{10.1002/ecs2.1345}.

\bibitem[Jaeger and Fahrig(2004)]{Jaeger2004}
Jochen A.~G. Jaeger and Lenore Fahrig.
\newblock Effects of {Road} {Fencing} on {Population} {Persistence}.
\newblock \emph{Conservation Biology}, 18\penalty0 (6):\penalty0 1651--1657,
  December 2004.
\newblock ISSN 0888-8892, 1523-1739.
\newblock \doi{10.1111/j.1523-1739.2004.00304.x}.
\newblock URL
  \url{https://conbio.onlinelibrary.wiley.com/doi/10.1111/j.1523-1739.2004.00304.x}.
\newblock Publisher: Wiley.

\bibitem[Kay and Giuggioli(2022)]{kay_diffusion_2022}
Toby Kay and Luca Giuggioli.
\newblock Diffusion through permeable interfaces: {Fundamental} equations and
  their application to first-passage and local time statistics.
\newblock \emph{Physical Review Research}, 4\penalty0 (3):\penalty0 L032039,
  September 2022.
\newblock \doi{10.1103/PhysRevResearch.4.L032039}.
\newblock URL \url{https://link.aps.org/doi/10.1103/PhysRevResearch.4.L032039}.
\newblock Publisher: American Physical Society.

\bibitem[Kay et~al.(2022)Kay, McKetterick, and Giuggioli]{kay_defect_2022}
Toby Kay, Thomas~John McKetterick, and Luca Giuggioli.
\newblock The defect technique for partially absorbing and reflecting
  boundaries: {Application} to the {Ornstein}–{Uhlenbeck} process.
\newblock \emph{International Journal of Modern Physics B}, 36\penalty0
  (07n08):\penalty0 2240011, March 2022.
\newblock ISSN 0217-9792.
\newblock \doi{10.1142/S0217979222400112}.
\newblock URL
  \url{https://www.worldscientific.com/doi/10.1142/S0217979222400112}.
\newblock Publisher: World Scientific Publishing Co.

\bibitem[Kenkre(2021)]{kenkre_montroll_2021}
V.~M. Kenkre.
\newblock The {Montroll} {Defect} {Technique} and {Its} {Application} to
  {Molecular} {Crystals}.
\newblock In V.~M. Kenkre, editor, \emph{Memory {Functions}, {Projection}
  {Operators}, and the {Defect} {Technique}: {Some} {Tools} of the {Trade} for
  the {Condensed} {Matter} {Physicist}}, Lecture {Notes} in {Physics}, pages
  213--243. Springer International Publishing, Cham, 2021.
\newblock ISBN 978-3-030-68667-3.
\newblock \doi{10.1007/978-3-030-68667-3_11}.
\newblock URL \url{https://doi.org/10.1007/978-3-030-68667-3_11}.

\bibitem[Kenkre and Sugaya(2014)]{kenkre_theory_2014}
V.~M. Kenkre and S.~Sugaya.
\newblock Theory of the transmission of infection in the spread of epidemics:
  interacting random walkers with and without confinement.
\newblock \emph{Bulletin of Mathematical Biology}, 76\penalty0 (12):\penalty0
  3016--3027, December 2014.
\newblock ISSN 1522-9602.
\newblock \doi{10.1007/s11538-014-0042-8}.

\bibitem[Langbein et~al.(2011)Langbein, Putman, and
  Pokorny]{langbein2011traffic}
Jochen Langbein, Rory Putman, and Bostjan Pokorny.
\newblock \emph{Traffic collisions involving deer and other ungulates in Europe
  and available measures for mitigation}, page 215–259.
\newblock Cambridge Usatniversity Press, 2011.
\newblock \doi{10.1017/CBO9780511974137.009}.

\bibitem[Lipton and Kaushansky(2018)]{lipton_first_2018}
Alexander Lipton and Vadim Kaushansky.
\newblock On the {First} {Hitting} {Time} {Density} of an
  {Ornstein}-{Uhlenbeck} {Process}.
\newblock \emph{arXiv}, 2018.
\newblock \doi{10.48550/arXiv.1810.02390}.

\bibitem[Lynch and Fagan(2009)]{lynch2009}
Heather~J Lynch and William~F Fagan.
\newblock Survivorship curves and their impact on the estimation of maximum
  population growth rates.
\newblock \emph{Ecology}, 90\penalty0 (4):\penalty0 1116--1124, 2009.
\newblock \doi{10.1890/08-0286.1}.

\bibitem[Martinez-Garcia et~al.(2020)Martinez-Garcia, Fleming, Seppelt, Fagan,
  and Calabrese]{martinez-garcia_how_2020}
Ricardo Martinez-Garcia, Christen~H. Fleming, Ralf Seppelt, William~F. Fagan,
  and Justin~M. Calabrese.
\newblock How range residency and long-range perception change encounter rates.
\newblock \emph{Journal of Theoretical Biology}, 498:\penalty0 110267, August
  2020.
\newblock ISSN 0022-5193.
\newblock \doi{10.1016/j.jtbi.2020.110267}.

\bibitem[Meijer et~al.(2018)Meijer, Huijbregts, Schotten, and
  Schipper]{meijer2018global}
Johan~R Meijer, Mark~AJ Huijbregts, Kees~CGJ Schotten, and Aafke~M Schipper.
\newblock Global patterns of current and future road infrastructure.
\newblock \emph{Environmental Research Letters}, 13\penalty0 (6):\penalty0
  064006, 2018.
\newblock \doi{10.1088/1748-9326/aabd42}.

\bibitem[Menezes et~al.(2025)Menezes, Calabrese, Fagan, Prado, and
  Martinez-Garcia]{Menezes2025}
Rafael Menezes, Justin~M. Calabrese, William~F. Fagan, Paulo~Inácio Prado, and
  Ricardo Martinez-Garcia.
\newblock The range-resident logistic model: a new framework to formalize the
  population-dynamics consequences of range residency.
\newblock \emph{Ecology Letters}, 28\penalty0 (12):\penalty0 e70269, 2025.
\newblock \doi{10.1101/2025.02.09.637279}.

\bibitem[Montroll and Potts(1955)]{Montroll1955}
Elliott~W Montroll and Renfrey~B Potts.
\newblock Effect of defects on lattice vibrations.
\newblock \emph{Physical Review}, 100\penalty0 (2):\penalty0 525, 1955.
\newblock \doi{10.1103/PhysRev.100.525}.

\bibitem[Moorcroft and Lewis(2013)]{moorcroft2013mechanistic}
Paul~R Moorcroft and Mark~A Lewis.
\newblock \emph{Mechanistic home range analysis.(MPB-43)}.
\newblock Princeton University Press, 2013.

\bibitem[Mueller and Fagan(2008)]{mueller2008search}
Thomas Mueller and William~F Fagan.
\newblock Search and navigation in dynamic environments--from individual
  behaviors to population distributions.
\newblock \emph{Oikos}, 117\penalty0 (5):\penalty0 654--664, 2008.
\newblock \doi{10.1111/j.0030-1299.2008.16291.x}.

\bibitem[Noonan et~al.(2022)Noonan, Ascensão, Yogui, and
  Desbiez]{noonan_roads_2022}
M.~J. Noonan, F.~Ascensão, D.~R. Yogui, and A.~L.~J. Desbiez.
\newblock Roads as ecological traps for giant anteaters.
\newblock \emph{Animal Conservation}, 25\penalty0 (2):\penalty0 182--194, 2022.
\newblock ISSN 1469-1795.
\newblock \doi{10.1111/acv.12728}.
\newblock URL \url{https://onlinelibrary.wiley.com/doi/abs/10.1111/acv.12728}.
\newblock \_eprint: https://onlinelibrary.wiley.com/doi/pdf/10.1111/acv.12728.

\bibitem[Noonan et~al.(2019)Noonan, Tucker, Fleming, Akre, Alberts, Ali,
  Altmann, Antunes, Belant, Beyer, Blaum, Böhning-Gaese, Cullen~Jr., de~Paula,
  Dekker, Drescher-Lehman, Farwig, Fichtel, Fischer, Ford, Goheen, Janssen,
  Jeltsch, Kauffman, Kappeler, Koch, LaPoint, Markham, Medici, Morato, Nathan,
  Oliveira-Santos, Olson, Patterson, Paviolo, Ramalho, Rösner, Schabo, Selva,
  Sergiel, Xavier~da Silva, Spiegel, Thompson, Ullmann, Zięba, Zwijacz-Kozica,
  Fagan, Mueller, and Calabrese]{noonan_comprehensive_2019}
Michael~J. Noonan, Marlee~A. Tucker, Christen~H. Fleming, Thomas~S. Akre,
  Susan~C. Alberts, Abdullahi~H. Ali, Jeanne Altmann, Pamela~Castro Antunes,
  Jerrold~L. Belant, Dean Beyer, Niels Blaum, Katrin Böhning-Gaese, Laury
  Cullen~Jr., Rogerio~Cunha de~Paula, Jasja Dekker, Jonathan Drescher-Lehman,
  Nina Farwig, Claudia Fichtel, Christina Fischer, Adam~T. Ford, Jacob~R.
  Goheen, René Janssen, Florian Jeltsch, Matthew Kauffman, Peter~M. Kappeler,
  Flávia Koch, Scott LaPoint, A.~Catherine Markham, Emilia~Patricia Medici,
  Ronaldo~G. Morato, Ran Nathan, Luiz Gustavo~R. Oliveira-Santos, Kirk~A.
  Olson, Bruce~D. Patterson, Agustin Paviolo, Emiliano~Esterci Ramalho, Sascha
  Rösner, Dana~G. Schabo, Nuria Selva, Agnieszka Sergiel, Marina Xavier~da
  Silva, Orr Spiegel, Peter Thompson, Wiebke Ullmann, Filip Zięba, Tomasz
  Zwijacz-Kozica, William~F. Fagan, Thomas Mueller, and Justin~M. Calabrese.
\newblock A comprehensive analysis of autocorrelation and bias in home range
  estimation.
\newblock \emph{Ecological Monographs}, 89\penalty0 (2):\penalty0 e01344, 2019.
\newblock ISSN 1557-7015.
\newblock \doi{10.1002/ecm.1344}.
\newblock URL \url{https://onlinelibrary.wiley.com/doi/abs/10.1002/ecm.1344}.
\newblock \_eprint: https://onlinelibrary.wiley.com/doi/pdf/10.1002/ecm.1344.

\bibitem[Noonan et~al.(2021)Noonan, Martinez-Garcia, Davis, Crofoot, Kays,
  Hirsch, Caillaud, Payne, Sih, Sinn, Spiegel, Fagan, Fleming, and
  Calabrese]{noonan_estimating_2021}
Michael~J. Noonan, Ricardo Martinez-Garcia, Grace~H. Davis, Margaret~C.
  Crofoot, Roland Kays, Ben~T. Hirsch, Damien Caillaud, Eric Payne, Andrew Sih,
  David~L. Sinn, Orr Spiegel, William~F. Fagan, Christen~H. Fleming, and
  Justin~M. Calabrese.
\newblock Estimating encounter location distributions from animal tracking
  data.
\newblock \emph{Methods in Ecology and Evolution}, 12\penalty0 (7):\penalty0
  1158--1173, 2021.
\newblock ISSN 2041-210X.
\newblock \doi{10.1111/2041-210X.13597}.
\newblock URL
  \url{https://onlinelibrary.wiley.com/doi/abs/10.1111/2041-210X.13597}.
\newblock \_eprint:
  https://onlinelibrary.wiley.com/doi/pdf/10.1111/2041-210X.13597.

\bibitem[O'Sullivan(2022)]{osullivan_moivre_2022}
Cormac O'Sullivan.
\newblock De {Moivre} and {Bell} polynomials, March 2022.
\newblock URL \url{http://arxiv.org/abs/2203.02868}.
\newblock arXiv:2203.02868 [math].

\bibitem[Pagany(2020)]{pagany_wildlife-vehicle_2020}
Raphaela Pagany.
\newblock Wildlife-vehicle collisions - {Influencing} factors, data collection
  and research methods.
\newblock \emph{Biological Conservation}, 251\penalty0 (August):\penalty0
  108758, 2020.
\newblock ISSN 00063207.
\newblock \doi{10.1016/j.biocon.2020.108758}.
\newblock URL \url{https://doi.org/10.1016/j.biocon.2020.108758}.
\newblock Publisher: Elsevier.

\bibitem[Pinto et~al.(2018)Pinto, Bager, Clevenger, and Grilo]{Pinto2018}
Fernando~A.S. Pinto, Alex Bager, Anthony~P. Clevenger, and Clara Grilo.
\newblock Giant anteater (\emph{Myrmecophaga tridactyla}) conservation in
  {Brazil}: {Analysing} the relative effects of fragmentation and mortality due
  to roads.
\newblock \emph{Biological Conservation}, 228:\penalty0 148--157, December
  2018.
\newblock ISSN 00063207.
\newblock \doi{10.1016/j.biocon.2018.10.023}.
\newblock URL
  \url{https://linkinghub.elsevier.com/retrieve/pii/S0006320718300910}.

\bibitem[Pitman and Yor(1997)]{pitman_lengths_1997}
James~W. Pitman and Marc Yor.
\newblock On the lengths of excursions of some {Markov} processes.
\newblock \emph{Séminaire de probabilités de Strasbourg}, 31:\penalty0
  272--286, 1997.
\newblock ISSN 2510-3660.
\newblock \doi{10.1007/BFb0119313}.
\newblock URL \url{http://www.numdam.org/item/SPS_1997__31__272_0/}.

\bibitem[Pynn and Pynn(2004)]{pynn2004moose}
Tania~P Pynn and Bruce~R Pynn.
\newblock Moose and other large animal wildlife vehicle collisions:
  implications for prevention and emergency care.
\newblock \emph{Journal of emergency nursing}, 30\penalty0 (6):\penalty0
  542--547, 2004.
\newblock \doi{10.1016/j.jen.2004.07.084}.

\bibitem[Queiroz and Gautam(1992)]{queiroz1992}
Cesar~AV Queiroz and Surhid Gautam.
\newblock \emph{Road infrastructure and economic development: some diagnostic
  indicators}, volume 921.
\newblock World Bank Publications, 1992.

\bibitem[Ricciardi and Sato(1988)]{ricciardi_first-passage-time_1988}
Luigi~M. Ricciardi and Shunsuke Sato.
\newblock First-{Passage}-{Time} {Density} and {Moments} of the
  {Ornstein}-{Uhlenbeck} {Process}.
\newblock \emph{Journal of Applied Probability}, 25\penalty0 (1):\penalty0
  43--57, 1988.
\newblock ISSN 0021-9002.
\newblock \doi{10.2307/3214232}.
\newblock URL \url{https://www.jstor.org/stable/3214232}.
\newblock Publisher: Applied Probability Trust.

\bibitem[Rowden et~al.(2008)Rowden, Steinhardt, and Sheehan]{rowden2008road}
Peter Rowden, Dale Steinhardt, and Mary Sheehan.
\newblock Road crashes involving animals in australia.
\newblock \emph{Accident Analysis \& Prevention}, 40\penalty0 (6):\penalty0
  1865--1871, 2008.
\newblock \doi{10.1016/j.aap.2008.08.002}.

\bibitem[Rytwinski and Fahrig(2015)]{Rytwinski2015}
Trina Rytwinski and Lenore Fahrig.
\newblock The impacts of roads and traffic on terrestrial animal populations.
\newblock \emph{Handbook of road ecology}, pages 237--246, 2015.
\newblock \doi{10.1002/9781118568170.ch28}.

\bibitem[Santos et~al.(2016)Santos, Santos, Santos-Reis, Pican{\c{c}}o~de
  Figueiredo, Bager, Aguiar, and Ascensao]{santos2016carcass}
Rodrigo Augusto~Lima Santos, Sara~M Santos, Margarida Santos-Reis, Almir
  Pican{\c{c}}o~de Figueiredo, Alex Bager, Ludmilla~MS Aguiar, and Fernando
  Ascensao.
\newblock Carcass persistence and detectability: reducing the uncertainty
  surrounding wildlife-vehicle collision surveys.
\newblock \emph{PloS one}, 11\penalty0 (11):\penalty0 e0165608, 2016.

\bibitem[Santos et~al.(2011)Santos, Carvalho, and Mira]{santos2011long}
Sara~M Santos, Filipe Carvalho, and Ant{\'o}nio Mira.
\newblock How long do the dead survive on the road? carcass persistence
  probability and implications for road-kill monitoring surveys.
\newblock \emph{PLos one}, 6\penalty0 (9):\penalty0 e25383, 2011.

\bibitem[Sato(1978)]{sato_moments_1978}
Shunsuke Sato.
\newblock On the moments of the firing interval of the diffusion approximated
  model neuron.
\newblock \emph{Mathematical Biosciences}, 39\penalty0 (1):\penalty0 53--70,
  May 1978.
\newblock ISSN 0025-5564.
\newblock \doi{10.1016/0025-5564(78)90027-5}.
\newblock URL
  \url{https://www.sciencedirect.com/science/article/pii/0025556478900275}.

\bibitem[Seigle‐Ferrand et~al.(2022)Seigle‐Ferrand, Marchand, Morellet,
  Gaillard, Hewison, Saïd, Chaval, Santacreu, Loison, Yannic, and
  Garel]{Seigleferrand2022}
Juliette Seigle‐Ferrand, Pascal Marchand, Nicolas Morellet, Jean‐Michel
  Gaillard, A.~J.~Mark Hewison, Sonia Saïd, Yannick Chaval, Hugo Santacreu,
  Anne Loison, Glenn Yannic, and Mathieu Garel.
\newblock On this side of the fence: {Functional} responses to linear landscape
  features shape the home range of large herbivores.
\newblock \emph{Journal of Animal Ecology}, 91\penalty0 (2):\penalty0 443--457,
  February 2022.
\newblock ISSN 0021-8790, 1365-2656.
\newblock \doi{10.1111/1365-2656.13633}.
\newblock URL
  \url{https://besjournals.onlinelibrary.wiley.com/doi/10.1111/1365-2656.13633}.

\bibitem[Siegert(1951)]{siegert_first_1951}
Arnold J.~F. Siegert.
\newblock On the {First} {Passage} {Time} {Probability} {Problem}.
\newblock \emph{Physical Review}, 81\penalty0 (4):\penalty0 617--623, February
  1951.
\newblock \doi{10.1103/PhysRev.81.617}.
\newblock URL \url{https://link.aps.org/doi/10.1103/PhysRev.81.617}.
\newblock Publisher: American Physical Society.

\bibitem[Silva et~al.(2021)Silva, Crane, and Savini]{silva2021road}
In{\^e}s Silva, Matt Crane, and Tommaso Savini.
\newblock The road less traveled: Addressing reproducibility and conservation
  priorities of wildlife-vehicle collision studies in tropical and subtropical
  regions.
\newblock \emph{Global Ecology and Conservation}, 27:\penalty0 e01584, 2021.
\newblock \doi{10.1016/j.gecco.2021.e01584}.

\bibitem[Silva and Calabrese(2024)]{silva_emerging_2024}
Inês Silva and Justin~M Calabrese.
\newblock Emerging opportunities for wildlife conservation with sustainable
  autonomous transportation.
\newblock \emph{Frontiers in Ecology and the Environment}, 22\penalty0
  (2):\penalty0 e2697, March 2024.
\newblock ISSN 1540-9295, 1540-9309.
\newblock \doi{10.1002/fee.2697}.
\newblock URL
  \url{https://esajournals.onlinelibrary.wiley.com/doi/10.1002/fee.2697}.

\bibitem[Slater(2002)]{slater_assessment_2002}
F.~M. Slater.
\newblock An assessment of wildlife road casualties – the potential
  discrepancy between numbers counted and numbers killed.
\newblock \emph{Web Ecology}, 3:\penalty0 33--42, 2002.

\bibitem[Smouse et~al.(2010)Smouse, Focardi, Moorcroft, Kie, Forester, and
  Morales]{smouse2010stochastic}
Peter~E Smouse, Stefano Focardi, Paul~R Moorcroft, John~G Kie, James~D
  Forester, and Juan~M Morales.
\newblock Stochastic modelling of animal movement.
\newblock \emph{Philosophical Transactions of the Royal Society B: Biological
  Sciences}, 365\penalty0 (1550):\penalty0 2201--2211, 2010.

\bibitem[Sullivan(2011)]{sullivan2011trends}
John~M Sullivan.
\newblock Trends and characteristics of animal-vehicle collisions in the united
  states.
\newblock \emph{Journal of safety research}, 42\penalty0 (1):\penalty0 9--16,
  2011.
\newblock \doi{10.1016/j.jsr.2010.11.002}.

\bibitem[Toral and Colet(2014)]{toral_stochastic_2014}
Raúl Toral and Pere Colet.
\newblock \emph{Stochastic {Numerical} {Methods}: {An} {Introduction} for
  {Students} and {Scientists}}.
\newblock Wiley-VCH, Weinheim, 1st edition edition, 2014.
\newblock ISBN 978-3-527-41149-8.
\newblock \doi{10.1002/9783527683147}.

\bibitem[Uhlenbeck and Ornstein(1930)]{uhlenbeck_theory_1930}
G.~E. Uhlenbeck and L.~S. Ornstein.
\newblock On the {Theory} of the {Brownian} {Motion}.
\newblock \emph{Physical Review}, 36\penalty0 (5):\penalty0 823--841, September
  1930.
\newblock \doi{10.1103/PhysRev.36.823}.
\newblock URL \url{https://link.aps.org/doi/10.1103/PhysRev.36.823}.
\newblock Publisher: American Physical Society.

\bibitem[van Langevelde and Jaarsma(2005)]{langevelde2005}
Frank van Langevelde and Catharinus~F Jaarsma.
\newblock Using traffic flow theory to model traffic mortality in mammals.
\newblock \emph{Landscape ecology}, 19:\penalty0 895--907, 2005.
\newblock \doi{10.1007/s10980-005-0464-7}.

\bibitem[Volkmer and Wood(2013)]{volkmer_note_2013}
Hans Volkmer and John~J. Wood.
\newblock A {Note} on the {Asymptotic} {Expansion} of {Generalized}
  {Hypergeometric} {Functions}.
\newblock \emph{Analysis and Applications}, December 2013.
\newblock \doi{10.1142/S0219530513500346}.
\newblock URL \url{https://www.worldscientific.com/worldscinet/aa}.
\newblock Publisher: World Scientific Publishing Company.

\end{thebibliography}
\end{document}